\documentclass[twocolumn]{aastex6}

\usepackage{xcolor}
\usepackage{appendix}
\usepackage{amsmath}
\usepackage{booktabs}
\newcommand{\cahk}{CaII H \& K}

\begin{document}
\shortauthors{Kado-Fong et al.}
\title{ Near-Infrared Spectroscopy of 5 Ultra-massive Galaxies at $1.7<\MakeLowercase{z}<2.7$}
\author{
Erin Kado-Fong \altaffilmark{1}, 
Danilo Marchesini\altaffilmark{1}, 
Z. Cemile Marsan\altaffilmark{1},
Adam Muzzin\altaffilmark{2},
Ryan Quadri\altaffilmark{3},
Gabriel Brammer\altaffilmark{4},
Rachel Bezanson\altaffilmark{5},
Ivo Labb\'e\altaffilmark{6},
Britt Lundgren\altaffilmark{7},
Gregory Rudnick\altaffilmark{8},
Mauro Stefanon\altaffilmark{6},
Tomer Tal\altaffilmark{9},
David Wake\altaffilmark{10,11},
Rik Williams\altaffilmark{12},
Katherine Whitaker\altaffilmark{13,14,15},
Pieter van Dokkum\altaffilmark{16}
}

\altaffiltext{1}{Department of Physics and Astronomy, Tufts University,
  Medford, MA 02155, USA}
\altaffiltext{2}{Department of Physics and Astronomy, York University, 4700 Keele St., Toronto, Ontario, Canada MJ3 1P3}
\altaffiltext{3}{George P. and Cynthia W. Mitchell Institute for Fundamental Physics and Astronomy, Department of Physics \& Astronomy, Texas A\&M University, College Station, TX 77843, USA}
\altaffiltext{4}{Space Telescope Science Institute, 3700 San Martin Drive, Baltimore, MD 21218, USA}
\altaffiltext{5}{Department of Astrophysics, Princeton University, Princeton, NJ, 08544, USA}
\altaffiltext{6}{Leiden Observatory, Leiden University, P.O. Box 9513, NL 2300 RA Leiden, The Netherlands}
\altaffiltext{7}{Department of Physics, University of North Carolina, Asheville, NC 28804, USA}
\altaffiltext{8}{Department of Physics and Astronomy, University of Kansas, Lawrence, KS 66045, USA}
\altaffiltext{9}{Department of Astronomy \& Astrophysics, University of California, Santa Cruz, CA, USA}
\altaffiltext{10}{Department of Astronomy, University of Wisconsin-Madison, 475 North Charter Street, Madison, Wisconsin 53706, USA}
\altaffiltext{11}{Department of Physical Sciences, The Open University, Milton Keynes MK7 6AA, UK}
\altaffiltext{12}{Uber Technologies Inc., 1455 Market St., 4th Floor, San Francisco CA 94103}
\altaffiltext{13}{Department of Astronomy, University of Massachusetts, Amherst, MA 01003, USA} 
\altaffiltext{14}{Department of Physics, University of Connecticut, Storrs, CT 06269, USA}
\altaffiltext{15}{Hubble Fellow}
\altaffiltext{16}{Department of Astronomy, Yale University, 260 Whitney Avenue, New Haven, CT 06511, USA}

\email{erin.fong@tufts.edu}
\date{\today}

\begin{abstract}
We present the results of a pilot near-infrared (NIR) spectroscopic campaign of five very massive
galaxies ($\log(\text{M}_\star/\text{M}_\odot)>11.45$) 
in the range of $1.7<z<2.7$. We measure an absorption feature redshift for one
galaxy at 
$z_\text{spec}=2.000\pm0.006$. For the remaining galaxies, 
we combine the photometry with the continuum from the spectra to
 estimate continuum redshifts and stellar population properties.
We define a continuum redshift ($z_{\rm cont}$ ) as one in which the redshift is estimated probabilistically
using EAZY from the combination of catalog photometry and the observed spectrum. 
We derive the uncertainties on the stellar population
synthesis properties using a Monte Carlo simulation and examine the
correlations between the parameters with and without the use of the spectrum
in the modeling of the spectral energy distributions (SEDs).
The spectroscopic constraints confirm the extreme stellar masses of the galaxies in our sample. 
We find that three out of five galaxies are quiescent (star formation rate
of $\lesssim 1 M_\odot~yr^{-1}$)
with low levels of
dust obscuration ($A_{\rm V} < 1$) , that one galaxy displays both high levels of
star formation and dust obscuration (${\rm SFR} \approx 300 M_\odot~{\rm yr}^{-1}$, $A_{\rm V} \approx 1.7$~mag), 
and that the remaining galaxy has properties 
that are intermediate between the quiescent and star-forming populations.

\end{abstract}


\section{Introduction}

The standard $\Lambda$CDM paradigm of structure formation depicts a universe 
dominated by dark matter. In this paradigm, dark matter structures form 
hierarchically, with low-mass haloes forming before their 
higher mass counterparts. It is therefore expected that galaxies would also
form in a hierarchical manner. However, archaeological studies of local 
galaxies show that today's most massive galaxies formed the bulk of their stars 
rapidly in the very early universe \citep[see, for example, ][]{thomas2005}: 
these observations are supported by the existence of a significant population 
of massive quiescent galaxies up to redshift $z \sim 3$.
\citep{ cimatti2002, kriek2006b, mancini2009, marchesini2010,brammer2011, marchesini2014, newman2015, hill2016}.

 Recent models of galaxy formation have been successful in addressing several long-standing
tensions between theoretical predictions of massive galaxy growth in the
early universe and observational results of such massive galaxies at high-$z$. 
These models predict that massive ellipticals
 will possess old, metal-rich stellar populations and the shortest formation 
timescales \citep{delucia2006, delucia2007}, but 
there are still persistent differences between observations
and theoretical predictions of the high-mass end of the stellar mass function.
These issues raise questions
about the evolution of massive galaxies:  
\cite{marchesini2014} propose a revised evolutionary path 
for the formation of today's ultra-massive 
galaxies (UMGs), in which the progenitor population at $1.5<z<2.5$ is dominated 
by heavily dust-obscured ($A_V \sim 2$~mag) massive star-forming galaxies, 
and only
$\sim40\%$ of the population is comprised of quiescent galaxies with 
relatively little dust obscuration.

\begin{deluxetable*}{ c c c c c c c c}
\tablewidth{0in}
\tablecaption{Pointing and observation information for the initial sample. \label{pointtable}}
\tablehead{
  \colhead{id} & \colhead{RA} & \colhead{DEC} & \colhead{Dates }& \colhead{Exp. Time (sec)} & Instrument \\
 & & & & \colhead{( exposure time x exposure count )} & } 
\startdata
COS-75355  & 10h02m28.49s & +02d02m13.70s &  12/18/13  &  600 x 10 = 6000 & GNIRS$^a$\\
COS-207144  & 10h00m33.48s & +02d28m54.74s &  03/07/13  &  600 x 18 = 10800 & GNIRS$^a$ \\
COS-90676  & 10h01m57.00s & +02d16m12.14s &   12/17/13  &  600 x 6 = 3600 & GNIRS$^a$\\
COS-189962$^c$  & 10h02m14.42s & +02d35m11.92s &  05/22/13, 05/23/13 &  600 x 18 = 10800 & GNIRS$^a$\\
COS-71929  & 10h01m40.60s & +01d58m57.47s &  05/04/13  &  908.8 x 4 = 3635.2 & FIRE$^b$ \\
COS-37207  & 09h59m42.59s & +01d55m01.55s &  05/03/13, 05/04/13  &  908.8 x 8 = 7270.4 & FIRE$^b$\\
\enddata

\tablecomments { 
  $^a$The Gemini Near-Infrared Spectrograph \citep{elias2006}. $^b$ The Folded-port InfraRed 
  Echellette \citep{simcoe2013}. $^c$ This galaxy was omitted from analysis due to poor observing 
  conditions.
}
\end{deluxetable*}

Though recent photometric campaigns have been highly successful in probing
the evolutionary path of UMGs, little is known about the evolution of the 
most massive galaxies (i.e., $\log{(M_{\star}/M_{\sun})}>11.5$) 
in the early universe.
According to 
measurements of the stellar mass function, the number density of such
galaxies appears constant from $z\sim4$ to $z\sim1.5$ 
\citep{marchesini2009,muzzin2013b}. In this stellar mass regime, model 
predictions are, at best, marginally consistent with these observations
\citep[][ but see also \citealp{henriques2015}]{vogelsberger2014, dave2016}.

Due to the rarity of these galaxies, however, a huge amount of data and manpower
are needed to generate a large enough sample to characterize these
populations. Very few high-$z$ galaxies in this mass regime have been confirmed
spectroscopically, and population studies rely almost entirely on photometry for
redshifts and stellar population characteristics.
Recent studies have furthered the spectroscopic effort 
for estimating the nature of highly massive galaxies at high redshift 
\citep{onodera2010, vandesande2011, onodera2012, bezanson2013, vandesande2013, belli2014, belli2015, marsan2015,marsan2016}, 
but the total sample of 
spectroscopically confirmed very massive, high-$z$ galaxies known to the 
community at the present is still not large enough to be able to 
leverage satisfactory statistical power in characterizing the underlying
distribution that govern the characteristics of such galaxies.


\begin{figure*}
\centering
\includegraphics[width=\linewidth]{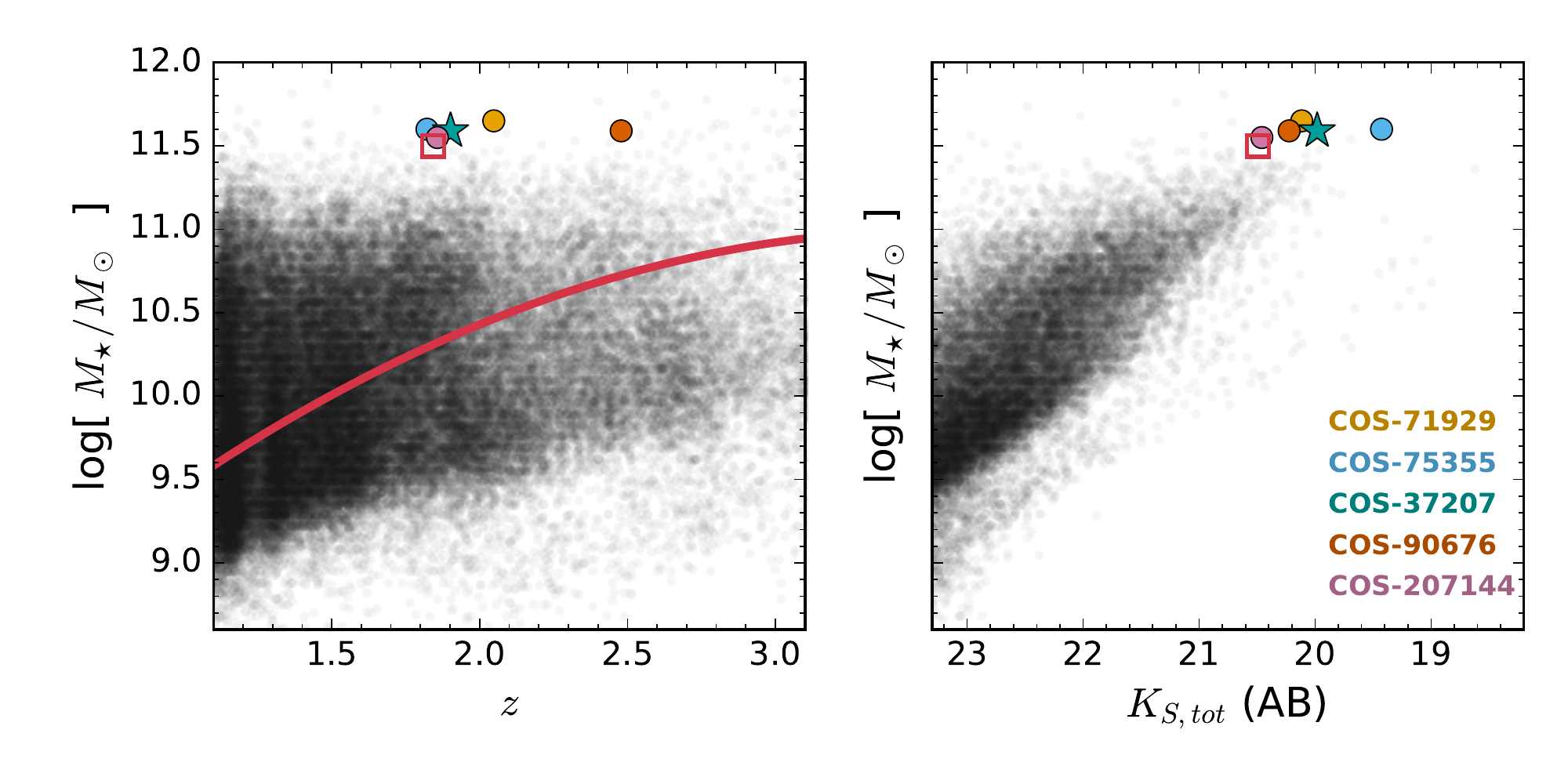}
\caption{\textit{Left:} Stellar mass versus redshift 
for the present sample (filled star and circles), COS-189962 (unfilled red square, not
included in sample due to poor observing conditions), and the UltraVISTA DR1 catalog of \cite{muzzin2013} as a 
whole (grayscale representation). 
The symbol type represents the classification of the galaxy as star-forming (star) or quiescent (circle)
based on the UVJ diagram (see \autoref{uvj}).
The red curve denotes the 90\% mass completeness
level for the catalog.
\textit{Right:} The same for stellar mass versus
 $K_s$ magnitude for galaxies at $1.7<z<2.7$. 
The sample at hand was selected to contain the brightest
and most massive galaxies in our targeted redshift range.}
\label{selectlmass}
\end{figure*}

In an ongoing effort to expand the sample of spectroscopically confirmed
very massive galaxies at high-$z$, we present the results of near-infrared (NIR) spectroscopy 
of 5 highly massive galaxies at $1.7 < z < 2.7$. 
In \autoref{sample}, we outline the sample selection process. The spectroscopic
observations are detailed in \autoref{spectroscopic_observations}. 
We derive robust redshifts and stellar population properties using a 
combination of the NIR spectroscopy and UltraVISTA 
photometry in \autoref{sedmodeling} in order to
confirm their status as ultra-massive, and to study the characteristics of 
this rare class of galaxies in the early universe. We present
spectroscopic redshifts and modeling results in \autoref{results}.
We place the characteristics derived for this sample of high-$z$ ultra-massive 
galaxies within the context of the population of such galaxies in 
\autoref{discussion_and_conclusions}.
In this work, we assume a \cite{chabrier2003} IMF, and adopt a 
cosmology of $\Omega_M = 0.3$, $\Omega_\Lambda = 0.7$, 
and $H_0$ = 70~km~s$^{-1}$~Mpc$^{-1}$.

\section{The Sample}\label{sample}

The targets of this study were chosen from the sample of ultra-massive galaxies
identified in the multiwavelength catalog constructed by \cite{muzzin2013}
across 1.62 deg$^{2}$ of the UltraVISTA/COSMOS field. The UltraVISTA
survey itself observes across four broadband 
filters \textit{Y,J,H} and 
\textit{Ks}, as well as one narrowband filter centered on $H\alpha$ at
z=0.8, which is not used in the present work \citep{mccracken2012}. 
The NIR coverage of UltraVISTA probes the rest-frame optical for galaxies
at $1.5<z<4.0$, an important range for the estimation of stellar population
properties.
The catalog from which the 
present sample was selected was constructed using the first data release of
the UltraVISTA survey; this data release spans about
one season of observing time, with a $5\sigma$ depth of approximately 
$K_s <23.9$ AB in a 2$^{\prime \prime}$ aperture.

The UltraVISTA catalog spans 30 filters from $0.15\mu m$ to 
$24.\mu m$, including ultra-violet (UV) imaging from the \textit{GALEX}
satellite \citep{martin2005}, and 
infrared coverage from \textit{Spitzer} \citep{sanders2007,frayer2009}. Sources are selected
from the $K_s$ band, which reaches
90\% completeness at $K_{s,TOT}=23.4$ AB, which corresponds to 
a stellar mass 90\% completeness
limit of $\log(\text{M}_\star/\text{M}_\odot)=10.43$ at $z=2$, as shown in
\autoref{selectlmass}. 
This broad wavelength coverage in the
UltraVISTA/COSMOS field allows us to confidently identify 
high-$z$ massive galaxies candidates for spectroscopic follow-up.
\autoref{selectlmass} shows the stellar mass ($M_{\star}$) 
versus redshift ($z$) and $K_s$ magnitude 
($K_s$,tot) versus $M_{\star}$ diagrams with the five galaxies targeted by 
our spectroscopic program highlighted as colored filled circles.

We originally selected all galaxies at $1.7<z<2.7$ with $\log{(M_{\star}/M_{\odot})}>􏰀11.5$, 
for a total of 18 candidates spanning the range in K$_{\rm S}$-band total magnitude of 
$19.4<K_{\rm S,tot}<23.0$. From this sample, we selected all galaxies (eight in total) 
brighter than $K_{\rm S,tot}=20.6$ to observe spectroscopically. 
Due to scheduling constraints, two of these galaxies were not observed. 
The exclusion of these two galaxies does not introduce any additional biases. Therefore, 
the six galaxies presented in our work constitute a representative sample of the 
population of brightest and most massive galaxies in the redshift range $1.5 \lesssim z \lesssim 3.0$.

\autoref{pointtable} gives the RA and DEC of each of the sources in our
sample along with observing and exposure information.

\section{Data}\label{spectroscopic_observations}
Four of the five galaxies in the sample (COS-75355, COS-90676, COS-207144, and COS-189962) 
were observed with the Gemini Near-infrared Spectrograph \citep[GNIRS,][]{elias2006}; 
the remaining
two galaxies were observed with the Folded-port InfraRed Echellette 
\citep[FIRE,][]{simcoe2013}
mounted on the Magellan 1 telescope. 
All of the galaxies in our sample were additionally imaged with HST WFC3 in the
F160W band (GO-12990; PI: Muzzin). 

\subsection{GNIRS}
Of the four galaxies observed with GNIRS, one (COS-189962) was discarded from the sample
due to poor seeing during the observations. 

GNIRS was set in its cross-dispersed mode with 32 line/mm grating and a 0.675" 
slit during the observing runs, which spanned 6 dates in March,
May, and December of 2013. 
The observations were made in queue mode, and are taken in 600
second exposures; the details of the observation runs are found in 
\autoref{pointtable}. The telluric emission lines present in the 
NIR place a limit on the per-exposure observation time; we therefore 
opt to take many shorter exposures and subsequently stack the resulting 
spectra in order to boost S/N.
The adopted GNIRS configuration provides a resolution of $R \approx 800$ 
and a wavelength range of around 0.8 to 2.4 $\micron$. A blind offset star 
and a B9V star were also observed for acquisition and to derive 
a telluric correction. 
Seeing averaged around 1.0$^{\prime \prime}$ across the observing run,
with minimal cloud cover (photometric conditions).

\subsection{FIRE Observations}

FIRE observations were made in its echellete mode with a 0.6$^{\prime \prime}$ slit and 
practical coverage from approximately $1.0\mu {\rm m}$ to $2.0 \mu {\rm m}$ with a spectral
resolution of $R\approx 6000$. 

Observations
were taken across two nights in May of 2013. 
As was the case
for the GNIRS observations, several short exposures were taken of each target
at 908.8 seconds per exposure due to the telluric emission lines present in 
the observed wavelength range. An A0V star was also observed in order
to perform telluric corrections.
Seeing averaged between 0.7$^{\prime \prime}$ and 1.0$^{\prime \prime}$ throughout the 
observing run, with minimal cloud cover. 
\\

\begin{table}
\begin{tabular}{lrlllll}
\toprule
&&\multicolumn{3}{c}{UltraVISTA DR1 ID}&& \\
\toprule
{} &  75355  &  207144 &  90676  &  189962 &  71929  &   37207  \\
filter       &         &         &         &         &         &          \\
\midrule
$m_{u^*}$  &    25.7 &   26.4  &   28.82 &   24.4  &   26.7  &   25.2   \\
$m_{g^{+}}$  &    24.8 &   26.3  &   26.8  &   24.3  &   26.1  &   24.7   \\
$m_{r^{+}}$  &    23.9 &   26.0  &   24.7  &   24.0  &   25.4  &   24.1   \\
$m_{i^{+}}$  &    23.0 &   25.1  &   24.2  &   23.7  &   24.4  &   23.5   \\
$m_{z^{+}}$  &    22.1 &   24.5  &   23.8  &   23.2  &   23.8  &   22.9   \\
$m_Y$        &    21.5 &   23.3  &   23.1  &   22.9  &   22.9  &   22.5   \\
$m_J$        &    20.3 &   22.1  &   22.1  &   21.7  &   21.7  &   21.3   \\
$m_H$        &    19.8 &   21.0  &   20.6  &   21.0  &   20.6  &   20.5   \\
$m_{K_s}$    &    19.4 &   20.5  &   20.2  &   20.5  &   20.1  &   20.0   \\
\bottomrule
\end{tabular}
\caption{Photometry from the UltraVISTA catalog for the six galaxies in the initial sample. 
  Total magnitudes are used.
  Magnitudes are given in the AB system. \label{magtable}}
\end{table}

\subsection{HST Imaging}
\begin{figure*}
\centering
\includegraphics[width=\linewidth]{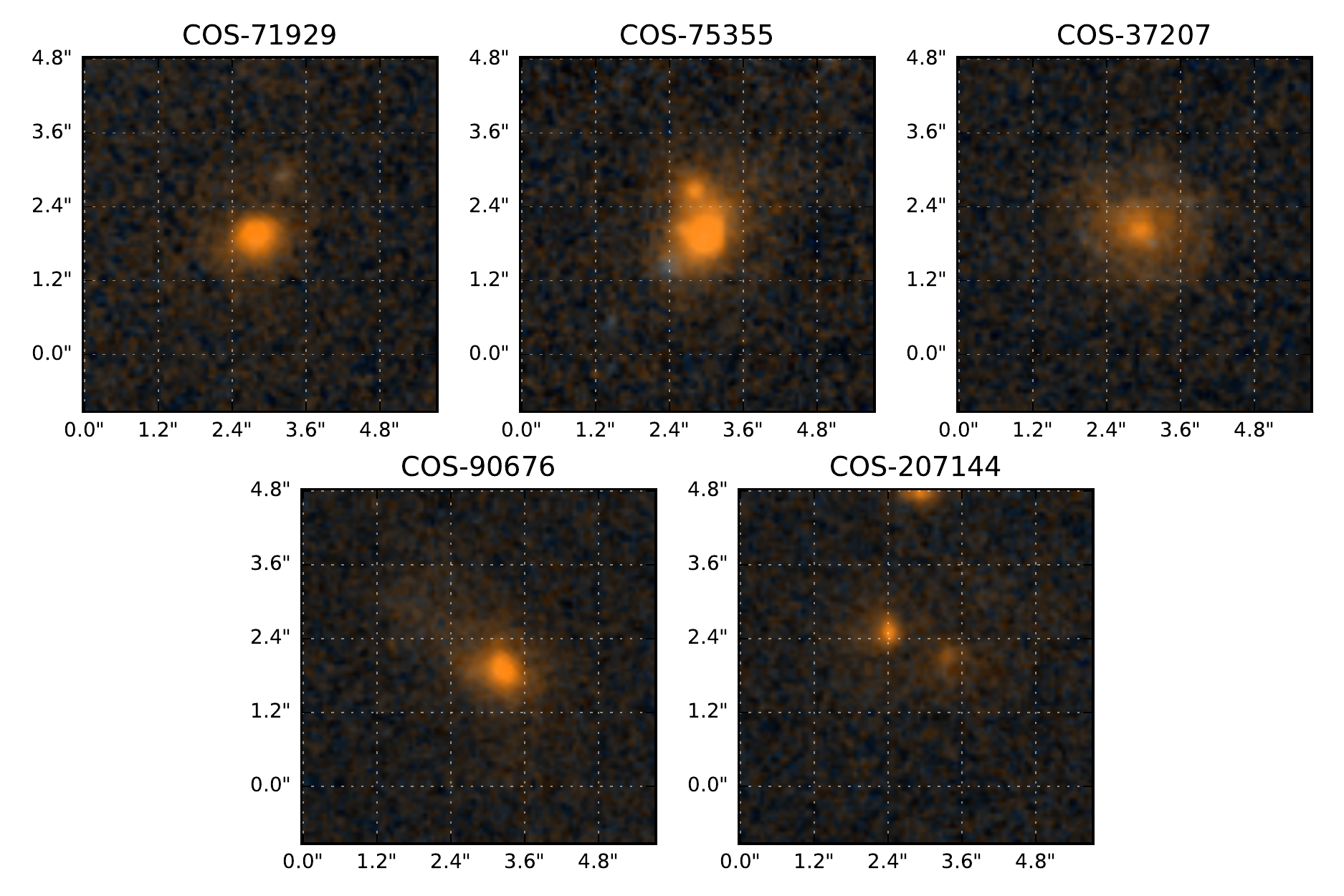}
\caption{Color cutouts of the galaxies in the present sample constructed from ACS I814 (F814W, blue) and 
  WFC3 H160 (F160W, red) images.}
\label{cutouts}
\end{figure*}

The galaxies at hand span a significant variety of morphologies,
as shown in \autoref{cutouts}:
COS-75355 appears to be consistent with a galaxy undergoing one to two minor 
mergers and COS-37207 is consistent with being a massive face-on red spiral, while
COS-71929 and COS-90676 are relatively isolated with some diffuse emission.
For a quantitative discussion of these morphologies, see Marsan et al. (2017; in prep.).
COS-207144, though identified as a single object in the UltraVISTA catalog,
is clearly resolved in two components in the HST image (see \autoref{cutouts}).
COS-207144 is the only object in our sample that falls within the 3D-HST survey 
\citep{skelton2014}. In 3D-HST, COS-207144 is resolved as two objects,
consistent with being an interacting pair at
$z=2.02^{+0.09}_{-0.14}$ and $z=2.07\pm 0.05$. There exists a grism redshift for
the latter object at $z_{\rm grism} = 2.362 \pm 0.015$ \citep{brammer2012,momcheva2016}. 
However, this grism redshift is based upon a low 
signal-to-noise, likely spurious, feature identified as $[OII]$ which,
if adopted, represents a shift of $>5\sigma$ from the original photometric 
redshift, implying that the two resolved objects are not physically
interacting. We therefore choose to compare the results of our analysis with 
those obtained using the original photometric redshift from \cite{skelton2014}.
 
\subsection{Data Reduction and Extraction of Spectra}

We performed the initial steps of the GNIRS data reduction 
(cosmic ray detection, bias subtraction, flat-fielding) using \textsf{pyraf}.
We then performed a first-pass sky subtraction by constructing a sky frame
using up to 4 dithered exposures taken at sufficiently similar times as the
target expsoure. 
We then ran a second pass sky subtraction by using the IRAF task 
\textsf{background} on the target exposure, which fits the sky on each side
of the galaxy continuum in order to remove remaining telluric emission lines
present in the frame.

To extract the spectra to one dimension, we used the spectrum of the 
telluric star as a trace for any remaining non-linearity in the spectrum shape. 
We then performed a telluric correction using the one dimensional spectrum 
of the observed telluric star. The orders (GNIRS's cross-dispersed mode 
yields six partial spectra corresponding to different wavelength ranges) were
then concatenated using a weighted average and flux calibrated using 
the UltraVISTA $H$ and $K_s$ photometry.

The IDL-based reduction pipeline FIREHOSE \citep{matejek2012} was used to reduce
the FIRE spectra. We adopted manual tracing for these 
spectra to account for the
faintness of the targets, but left the pipeline otherwise intact.

In order to increase the signal-to-noise ratios of these observations, we 
binned each spectrum with bins of variable size. 
As a result, bin size is not constant across a given spectrum -- areas
of low signal-to-noise are binned more as compared to areas in which
the signal-to-noise was relatively high in the unbinned spectrum.
Areas of low transmission, strong skyline residuals, or artifacts from
remaining skyline residuals  were masked. The resulting 
bin size (observed frame)
ranges from 51\AA~($15^{\rm th}$ percentile) to 163\AA~($85^{\rm th}$ percentile),
with a median bin size of 77\AA.
The resulting signal-to-noise ratio ranges from 2.51 ($15^{\rm th}$ percentile)
to 13.34 ($85^{\rm th}$ percentile), with a median signal-to-noise ratio of 6.99.

\begin{figure*}
\centering
\includegraphics[width=0.95\linewidth]{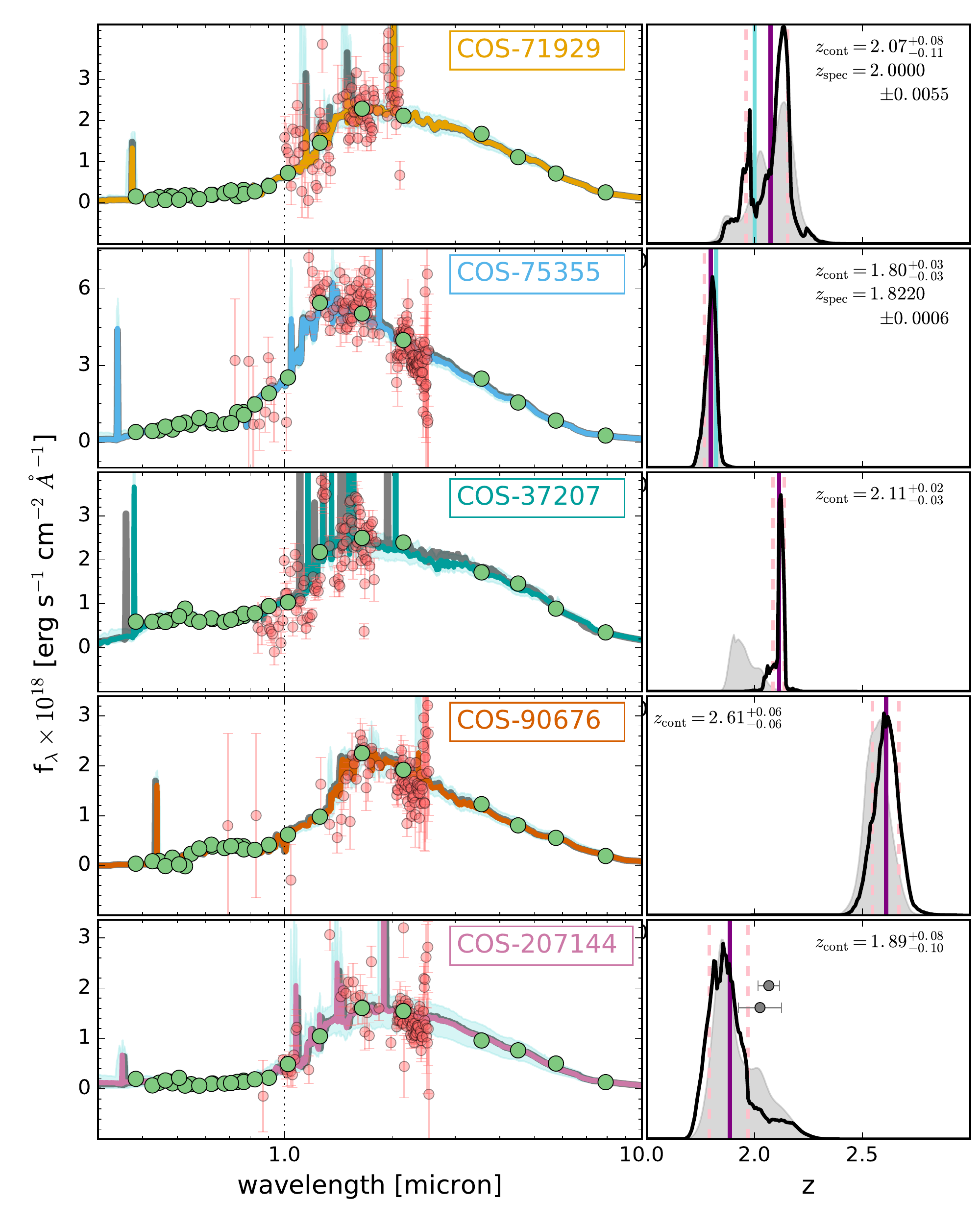}
\caption{\textit{Left:} Binned spectra for each galaxy (red filled circles)
 with the best-fit
EAZY model (color corresponds to the galaxy's label in each panel). 
Fluxes are scaled by a factor of $10^{18}$.
The light blue shaded region encases the upper and lower envelopes of the 
EAZY models that correspond to redshifts within the 68\% confidence bound of the redshift
posterior distribution. 
The green filled circles show the UltraVISTA catalog photometry.
\textit{Right:}EAZY redshift posterior distributions. The black curve
shows the distribution estimated using both the spectrum and photometry. The
grey shaded region shows the same distribution from the photometry alone.
$z_\text{cont}$ values are shown as purple lines, with the 68\% confidence
intervals marked with dashed pink lines. The $z_\text{spec}$ of
COS-75355 from \cite{onodera2012} and of COS-71929 from this work are shown as
 cyan vertical lines. The grey circles in the lower right panel denote the 
photometric redshifts derived for the two resolved objects that make up 
COS-207144 in \cite{skelton2014}.}
\label{allspec}
\end{figure*}

\autoref{allspec} shows the final binned 1D spectra of the five sources
(red filled circles), overplotted with the UltraVISTA broadband and 
medium-band photometry (green filled circles).

\section{SED Modeling}\label{sedmodeling}

We combined the binned spectra with the
UltraVISTA photometry to model the spectral energy distributions (SEDs) of
galaxies using the photometric redshift
code EAZY \citep{brammer2008} and the stellar population synthesis code FAST
\citep[Fitting and Assessment of Stellar Templates,][]{kriek2009}. 
The addition of the spectra allows us to derive continuum redshifts 
and stellar population properties from the two codes
by taking advantage of the
coverage provided by the broadband photometry and the finer observation 
of the 4000\AA ~break delivered by the spectroscopy, producing robust 
redshift estimates in the absence of emission or absorption features. 

Following Muzzin et al. (2013b), EAZY fits a linear combination of 
six templates from the PEGASE models \citep{fioc1997}, a 
red template from the model library of \cite{maraston2005}, 
a $\sim$1 Gyr old post-starburst template, as well as a 
slightly dust-reddened Lyman Break template.
EAZY also incorporates a
template error function to account for larger uncertainties in the models in the
rest-frame NIR relative to the rest-frame optical. A detailed description of
this process may be found in \cite{brammer2008}. EAZY was run using the v1.0
template error function and a K-band magnitude prior.

We used FAST to estimate stellar population properties.
FAST produces best-fit parameters for stellar mass, star formation rate, 
stellar age, characteristic timescale ($\tau$) of the assumed
star formation history (i.e., delayed exponentially declining
star formation history), extinction (using the 
extinction curve detailed in \citealt{kriek2013}) and metallicity. In this 
analysis we set an allowed stellar age range of
 $\log{(\text{age}~[\text{yr}]~ )}=8-10.1$, an 
allowed range of $\tau$ of $\log{(\tau~ [\text{yr}]~ )}=7-10$, 
and an allowed extinction
 range of $A_{\rm V}=0-5$~mag.”
We used the high-resolution version of the flexible stellar population synthesis (FSPS)
 models from \cite{conroy2009,conroy2010b}
with a \cite{chabrier2003} IMF. 

In order to estimate errors on the stellar population properties derived from FAST, we 
constructed a Monte Carlo (MC) simulation external to FAST and EAZY. The input 
photometry and spectra were perturbed 
according to a normal distribution described by $\sim~N(f_i, \sigma_i)$,
where $f_i$ is the mean flux of the $i^\text{th}$ measurement, 
and $\sigma_i$ is the uncertainty of the $i^\text{th}$ measurement
in the spectrum and photometry. At this juncture, we 
implicitly assumed that each wavelength bin is uncorrelated from its neighbors; 
we do expect there to be some degree of correlation between adjacent flux bins in
the spectra, but the
impact of a bin-to-bin correlation will be an overestimation of the MC-derived
errors. We therefore chose to assume that each bin is fully independent of
the state of nearby wavelength bins.

For those galaxies that
have spectroscopic redshifts, the FAST redshift was held fixed for all runs.
For the galaxies that do not have spectroscopic redshifts, 
we run EAZY using the perturbed spectrum and photometry, and run 
FAST with the perturbed data, holding the redshift fixed to the
best-fit redshift of the corresponding EAZY output.

This method of estimation allows for the independent derivation of errors on 
the stellar population properties estimated by FAST and EAZY. For all galaxies, 
500 simulations were run for both FAST and EAZY. A full description of the
parameter correlations that result from these simulations is given in
\autoref{Appendix}.

\begin{figure}
\centering
\includegraphics[width=\linewidth]{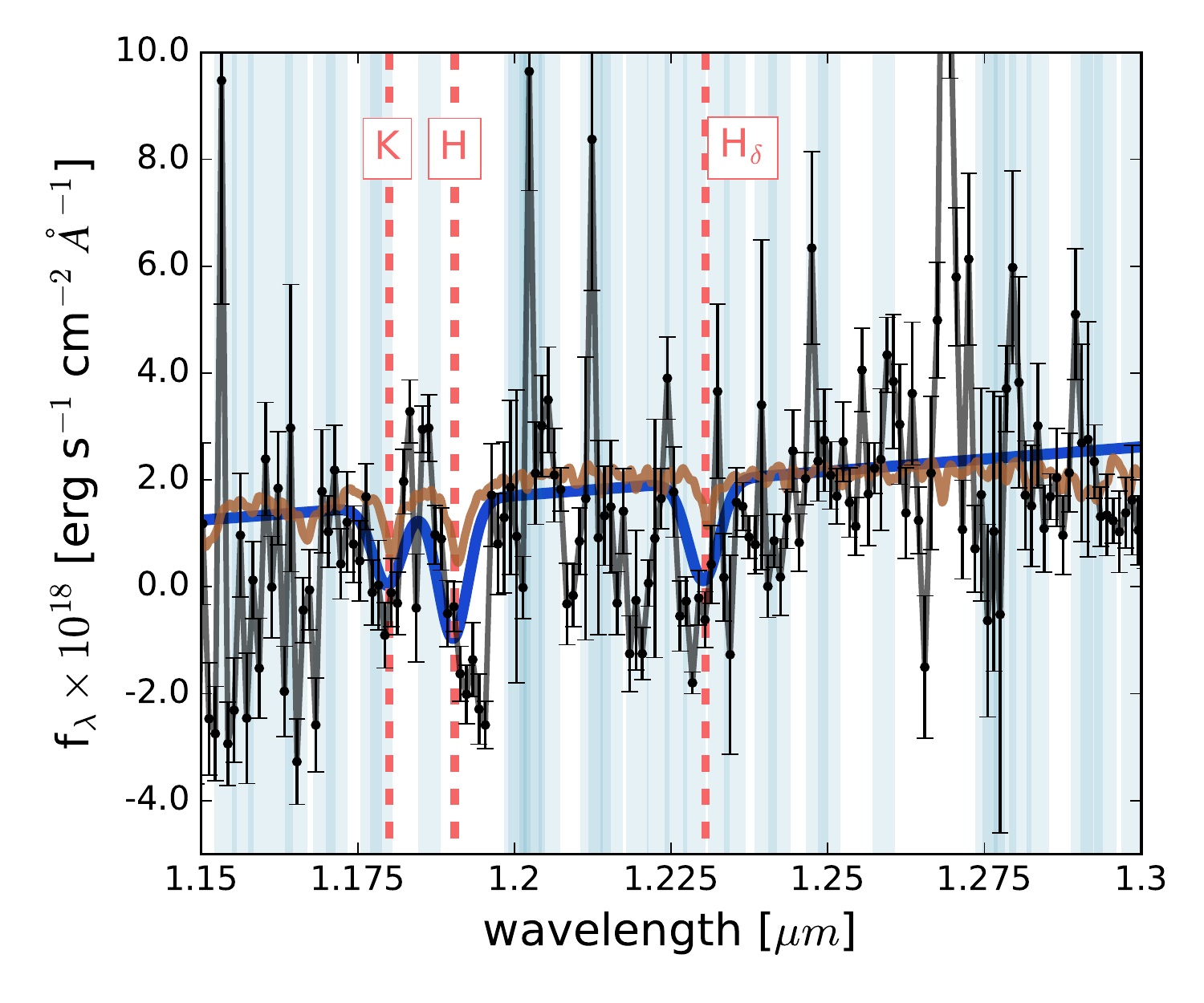}
\caption{\cahk{} absorption features in COS-71929.  
The brown curve
shows the best-fit EAZY SED to the galaxy at $z=2.000$. The red dashed lines
show the placement of the \cahk{} and $H_\delta$ features, as labeled, at $z=2.000$.
The blue curve shows the fit to a triple
Gaussian with the standard deviations and redshifts constrained to be equal
for all features. The light blue shaded regions show areas in which OH 
skylines were removed during data reduction. }
\label{cahk}
\end{figure}

\section{Results}\label{results}

\subsection{Spectroscopic Redshifts}

There are spectroscopic redshifts for 2 galaxies in the sample.
First, COS-75355 has a literature $z_\text{spec}=1.822$ from \cite{onodera2012}.
Secondly, the spectrum of COS-71929 shows absorption features consistent with 
\cahk{}, allowing us to measure a spectroscopic redshift
 at $z_\text{spec}=2.000\pm0.006$. 

\cite{onodera2012} finds a spectroscopic redshift for COS-75355 using the 
\cahk{} absorption features, in conjunction with $\text{H}_\beta$ in 
absorption. The spectrum in our sample does not have sufficient
signal-to-noise to detect most of the
features used to identify the spectroscopic redshift -- there is a possible
detection of Na in absorption at $z=1.822$, but the detection is not
secure.

The features used to determine a spectroscopic redshift for COS-71929 are
shown in \autoref{cahk}, as labeled. The best-fit EAZY model SED
is overlaid in brown. From this spectrum we constrained our spectroscopic
redshift for COS-71929 to be $z_{\rm spec}=2.0000 \pm 0.0055$, which is within one sigma of the 
continuum redshift, as detailed in the following section.

We simultaneously fit the \cahk{} and H$_\delta$ features using 
a triple Gaussian with the
observed velocity dispersion and redshift constrained to be equal for all features.
The underlying continuum was approximated as linear in this regime.
From this treatment, we derived a formal error for the spectroscopic redshift
of $\pm 0.0055$. We note that the spectroscopic redshift for COS-71929 is
located at the secondary peak in the galaxy's
redshift probability density function, with the bimodality
of the probability density function likely to be caused by a degeneracy between the locations
of $\rm{CaII}~H$ and H$_\delta$.

We additionally attempted to approximate a constraint on the stellar velocity dispersion
of COS-71929 using the fit detailed above. The instrumental spectral resolution
is given to be 50~km~s$^{-1}$; we subtracted this value in quadrature from the observed
velocity dispersion to make an estimate of the stellar velocity dispersion.
We estimated the uncertainty in the fit due to 
measurement error by running a Monte Carlo simulation to convergence in which
the binned flux values were perturbed according to a normal distribution 
with a standard deviation equal to the measurement uncertainty. This procedure
gives a velocity dispersion of $549^{+106}_{-262}$~km~s$^{-1}$, where the
quoted errors are the 2-sided 95\% confidence bounds. 
However, we stress that our
measurement of the velocity dispersion is very crude.
The low SNR of the spectrum does not allow for a proper
analysis of the velocity dispersion, 
and our measurement is subject to extremely high
uncertainty, with a 1-sided 99\% lower confidence bound of 217~km~s$^{-1}$.

\subsection{Continuum Redshifts}
\begin{figure}
\centering
\includegraphics[width=\linewidth]{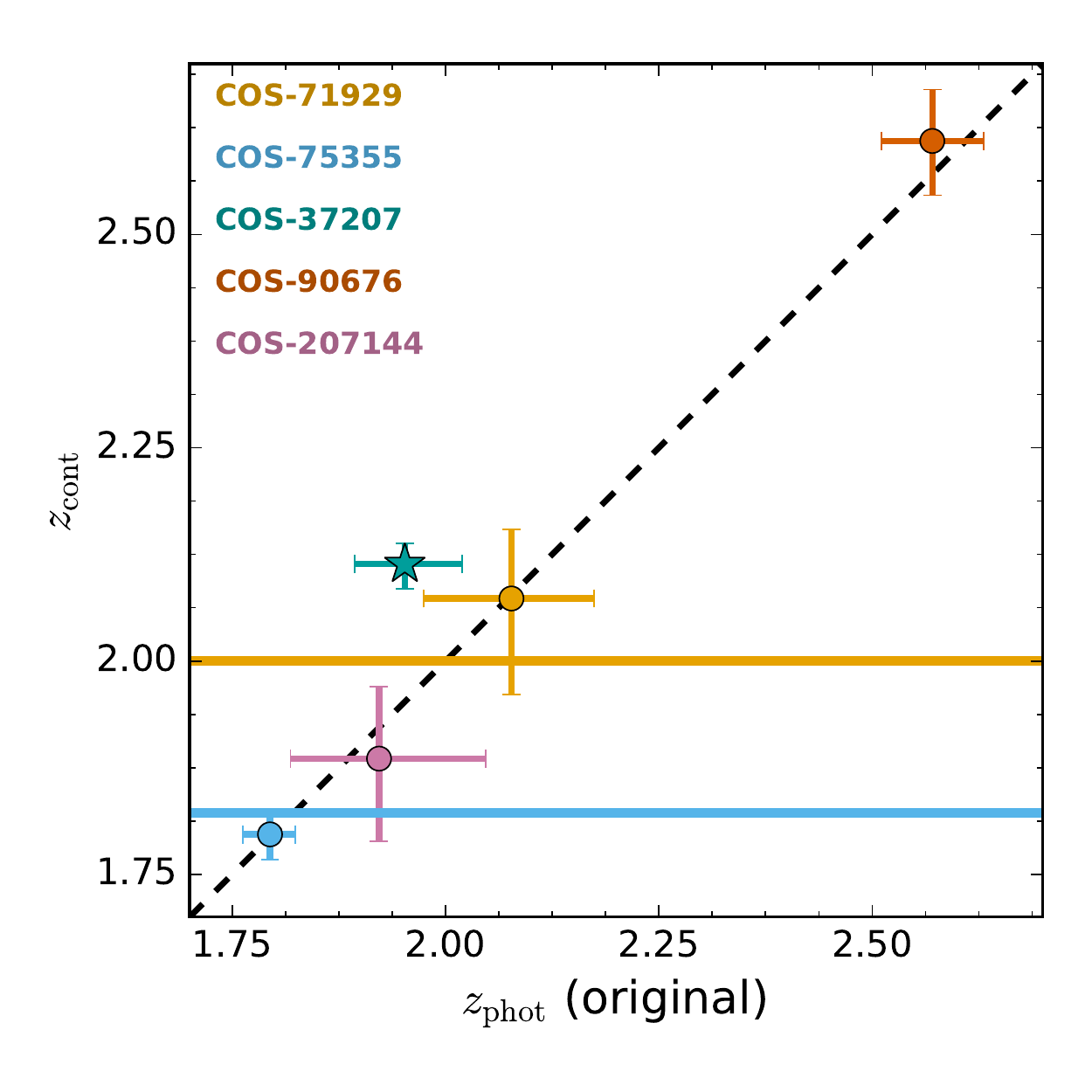}

\caption{Continuum EAZY redshifts ($z_{\rm cont}$) versus photometry-only EAZY redshifts ($z_{\rm phot}$).
Spectroscopic redshifts are shown for COS-75355 and COS-71929 in blue and yellow
horizontal lines, respectively. Quiescent galaxies are marked as circles; the 
star-forming galaxy COS-37207 is shown as a star.
The black dashed line shows the 1:1 relation. }

\label{zphotzspec}
\end{figure}

For all of the galaxies in the sample (including those that have spectroscopic
redshifts), we estimated a continuum redshift with EAZY ($z_{\rm cont}$), 
in which both
the photometry and spectroscopy are used in the fit. 
\autoref{zphotzspec} shows the continuum redshifts, $z_{\rm cont}$, for the
galaxies in our sample against their original photometric redshifts ($z_{\rm phot}$, derived
from the UltraVISTA DR1 photometry only). The quoted errors in  
\autoref{zphotzspec} are the 68\% confidence intervals from the EAZY
redshift probability density function.

The continuum redshifts of both COS-75355 and COS-71929 are in good
agreement ($<1\sigma$ difference) with their spectroscopic redshifts.
The spectroscopic redshifts of both galaxies are shown as horizontal lines in
\autoref{zphotzspec}. Although the errorbars shown on the figure are taken
from the EAZY redshift probability density function (PDF), 
the distribution of our EAZY Monte Carlo simulation
are in good agreement with the output EAZY PDF. The
main peak of the EAZY redshift probability density function is somewhat higher
than the spectroscopic redshift for COS-71929 (see \autoref{allspec}),
but there is a secondary peak in the EAZY redshift PDF that is consistent with
our spectroscopic redshift. Furthermore, the spectroscopic redshift is within
$1\sigma$ of the continuum redshift for the galaxy.

The formal errors on the continuum redshift are lower than the original
photometric redshift in all cases when
the spectrum is added to the fit, though COS-37207 is the only galaxy 
whose continuum redshift is significantly different than the 
original photometric redshift. 
Although the original photometric redshift of COS-37207 
lies outside of the continuum redshift 95\% confidence interval, 
the absolute difference between
the continuum and photometric redshifts is relatively small
($\sim0.16$; i.e., $\Delta z/(1+z) \sim 0.05$”).

As the photometric redshifts for both of the galaxies that make up
COS-207144 are available, it is informative to compare our results against the
photometric redshifts available from \citep{skelton2014}. Given that we 
expect the majority of the light in the observed spectrum to originate from the
brighter galaxy (which, in the NIR, is the galaxy at $z=2.02$), it is unsurprising
that our continuum redshift is in better agreement with the brighter of
the two objects ($<1\sigma$ difference) than with the fainter ($<2\sigma$ difference).

\begin{figure}
\includegraphics[width=\linewidth]{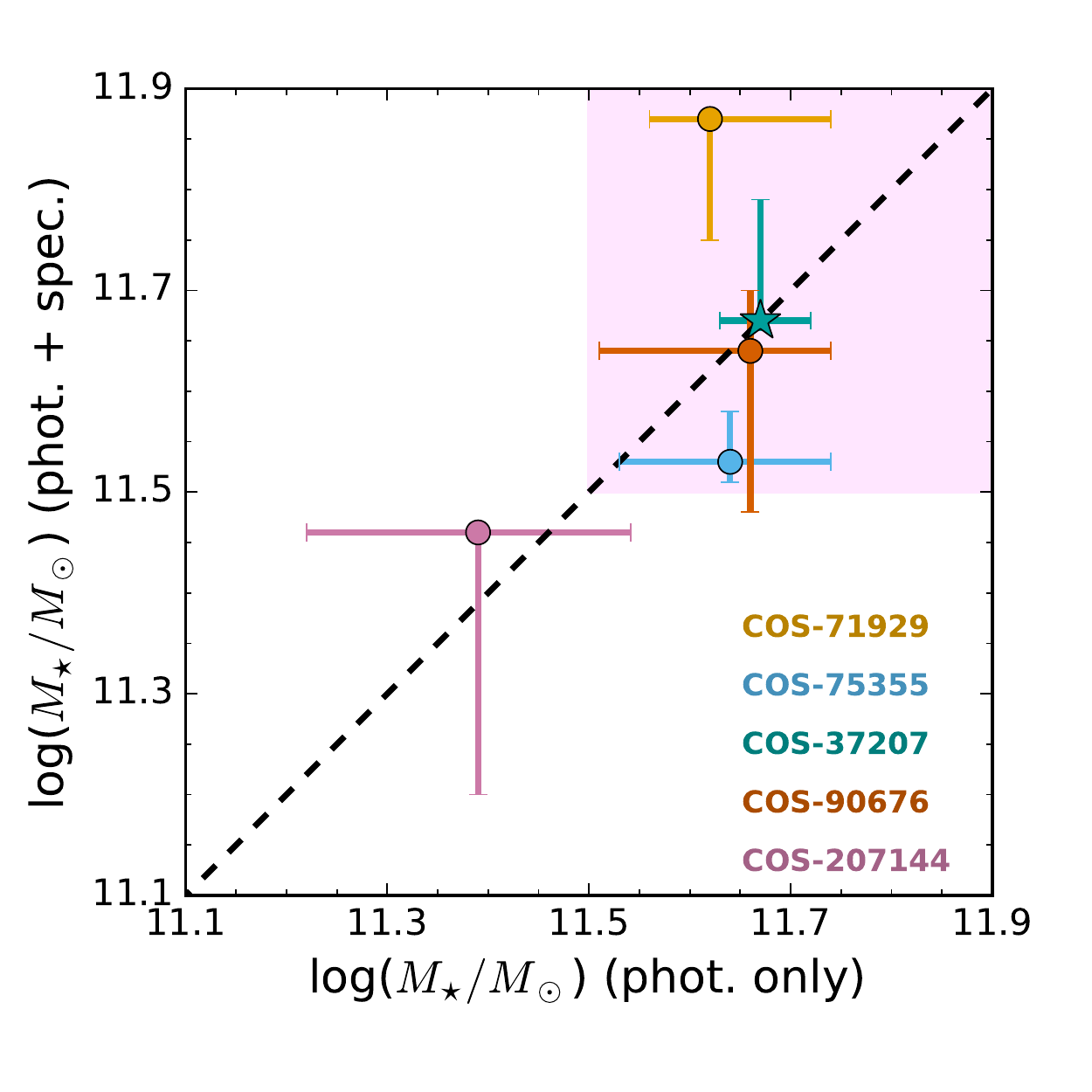}
\centering
\caption{Continuum FAST stellar mass estimates versus photometry-only
 FAST stellar mass estimates. The black dashed line shows 
the 1:1 relation. The pink shaded region marks stellar masses
where log($\text{M}_\star/\text{M}_\odot$)$>11.5$. As in \autoref{zphotzspec},
quiescent galaxies are shown with circles and the star-forming galaxy with a star.}
\label{photspecmass}
\end{figure}

\subsection{Stellar Population Properties}
We used the procedure outlined in \autoref{sedmodeling} to estimate 
the stellar population properties of the sample. \autoref{fastoutput} shows
the FAST output for the five galaxies, along with their 1$\sigma$ errors.
The best-fit values that we quote are the original FAST best-fit values,
and not average values of the Monte Carlo distributions. Thus, in some cases,
the best-fit value lies on the edge of the 1$\sigma$ interval. Re-defining our
best-fit values by using an average of the Monte Carlo distributions does not
qualitatively change our results.

We confirm that all of the galaxies in our sample are indeed ultra-massive.
All galaxies in the sample 
have a stellar mass exceeding $\log(\text{M}_\star/\text{M}_\odot)\gtrsim11.5$. 
In the case of COS-207144 (the blended object), stellar mass estimates obtained
here may be compared against those obtained for the objects resolved in 
3DHST by \cite{skelton2014}. The best-fit stellar mass of the brighter 
galaxy of the pair is 
in good agreement (within 68\% confidence interval) with that of COS-207144. 
The sum of the best-fit stellar masses of the resolved objects 
is significantly larger ($\log ( M_*/M_\odot )=11.62$) than our best-fit mass for
COS-207144. This is likely a result of the brighter galaxy dominating the 
light of the spectrum -- the best-fit stellar mass of COS-207144 
($\log(M_*/M_\odot)=11.46 ^{+0.00}_{-0.26}$) 
is in strong agreement with the best-fit stellar mass of this source
as derived by \cite{skelton2014} of
$\log (M_*/M_\odot)=11.40$.

Given that the original selection criteria from the UltraVISTA DR1 
catalog was based partially on stellar mass, we plot the
stellar mass FAST estimated from photometry alone against the same quantity from
the photometry in conjunction with the spectroscopy in \autoref{photspecmass}.
None of the updated best-fit stellar masses differ significantly from the
stellar mass derived from the photometry alone.
In the case
of COS-37207, the galaxy for which the best-fit redshift changed the most
significantly when the spectrum was added to the fit (see \autoref{allspec}), 
the formal error on the stellar mass increased when the spectrum is
added, with relatively little change to the best-fit stellar mass.

As a probe of the evolutionary
stage of the galaxies in our sample, we computed the quenching factor 
$q_\text{sf}$ \citep{kriek2009} , which is $q_{\rm sf}=1-b$, where $b$ is the birthrate
parameter \citep{kennicutt1983, scalo1986}. The quenching factor is therefore

\begin{equation}
q_\text{sf} = 1 - \frac{\text{SFR}_\text{current}}{\langle \text{SFR}_\text{past}\rangle} = 1~-~\frac{\text{SFR}_\text{current}}{M_\star / \text{age} },
\end{equation}

such that a fully quenched galaxy will have a quenching factor of 1.
With the exception of COS-37207, which has a quenching factor of 
 $0.276^{+0.529}_{-0.276}$, all of the galaxies in our sample have quenching factors
consistent with $q_\text{sf} > 0.90$. 

COS-37207 is the only galaxy in the sample to show high levels of star 
formation, with a best-fit SED star formation rate (SFR) 
of $339^{+108}_{-221} M_\odot ~ {\rm yr}^{-1}$. The error bars quoted here 
correspond to the 68\% confidence interval. The 99\% confidence interval bounds
are [31,788] $M_\odot ~ {\rm yr}^{-1}$, the lower bound of which is greater than the
best-fit SFR for any of the other galaxies in the sample. 
We therefore conclude that this galaxy is
securely identified as star forming.
In light of the fact that this galaxy has a high estimated
SFR, but is also characterized by a large dust extinction of $A_V=1.7^{+0.1}_{-0.4}$, 
it is of interest to note that the spectrum of COS-37207 does not 
show any secure emission lines. Using the best-fit values of SFR and $A_V$,
we estimated the expected observed flux for $[OII]$, $H_\beta$, and $[OIII]$
using the relations from \cite{calzetti2000} and \cite{price2014}. From these
estimates, we conclude that neither $H_\beta$ nor $[OII]$ would be observable
given the noise properties of the spectrum. $[OIII]$ may have been visible, but
coincides with a strong skyline at the best-fit redshift of $z=2.11$. 
It is therefore expected that strong emission lines are absent
from the spectrum of COS-37207. 

Though COS-71929 also shows evidence of some star formation 
(SFR~$=8.7^{+0.4}_{-6.9} M_\odot~ {\rm yr}^{-1}$),
a measurement of the quenching factor of this galaxy indicates 
that the SFR was significantly higher in the past ($q_\text{sf} \approx 0.95$).

Additionally, the estimated dust extinction for COS-37207 is the 
highest in the sample, with $A_V=1.70^{+0.10}_{-0.40}$ mag.  COS-71929 displays
the second-highest level of dust extinction at  $A_V = 1.30^{+0.10}_{-0.40}$ mag.
We therefore characterize COS-37207 as a dusty star-forming galaxy,
COS-71929 as a galaxy with mostly suppressed star-formation intermediate between
star-froming and quiescent galaxies, and the rest of the galaxies in our
sample as quiescent. 


\begin{figure}
\centering
\includegraphics[width=\linewidth]{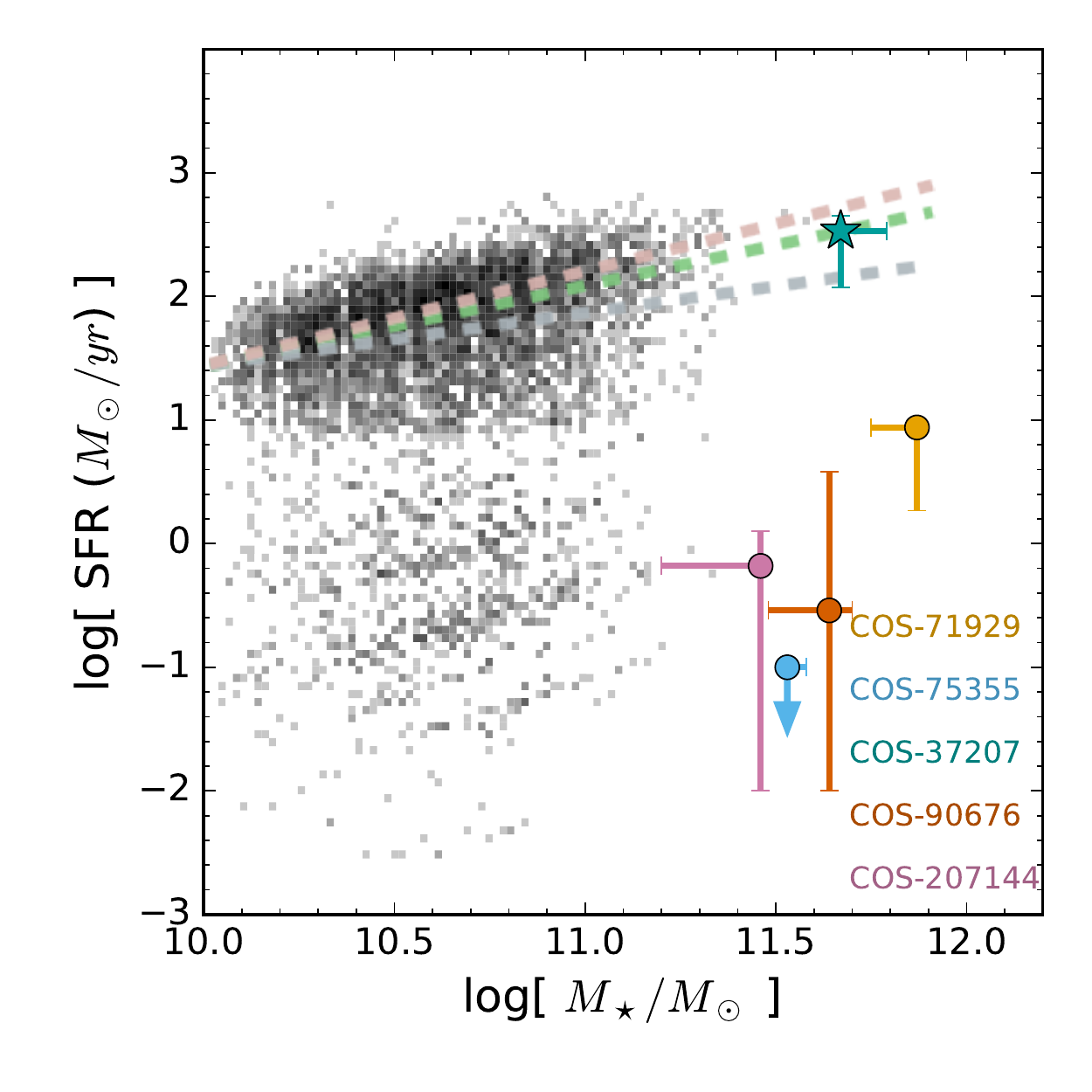}
\caption{Stellar mass versus star formation rate for the sample (colored points) and the
mass-complete UltraVISTA catalog in the range of $1.7<z<2.7$. The average star-forming
main sequence from \cite{whitaker2014} for the redshift range in question is shown by the
green dashed line. 
The parameterization of the star-forming main sequence by \cite{tomczak2016} is shown for
$z=2.15$ by the blue dashed line. The parameterization of the star-forming main sequence as
derived by \cite{speagle2014} at $t=3.0$~Gyr is shown by the red dashed line.
As expected,
the four quiescent galaxies in the sample (filled circles) lie below the star-forming main sequence,
whereas COS-37207 (the dusty star-forming galaxy, filled star) lies along the star-forming main sequence. }
\label{sfrmass}
\end{figure}

We plot our stellar mass and SFR estimates against the mass-complete sample of
galaxies at $1.7<z<2.7$ from the UltraVISTA catalog in \autoref{sfrmass}, 
along with the star-forming main sequences measured by \cite{speagle2014}, \cite{whitaker2014},
and \cite{tomczak2016}. For the UltraVISTA catalog, we preferentially use UV+IR SFR measurements. In the
abscence of such a measurement (or a non-detection), we use the SED SFR.
The FAST-derived stellar masses and SFRs for our sample place the three
quiescent galaxies (COS-75355, COS-207144, and COS-90676) on the high-mass end
of the quiescent population. COS-37207 lies along the high-mass
end of the star-forming main sequence, making it the most massive star-forming galaxy 
in the targeted redshift range.
Finally, COS-71929, with a SFR$\approx$10~M$_{\odot}$~yr$^{-1}$ is 
seen to lie well below the main sequence of star-forming galaxies, indicating that the
 star formation activity of COS-71929 is mostly quenched.
 We note that this picture qualitatively holds
 if the SED-modeling SFRs are replaced with the SFR from the combination of ultra-violet 
(i.e., unobscured) and infrared (dust reprocessed) estimates of the SFRs (see Section 5.5).

All of the galaxies in our sample have nominal best-fit stellar ages greater than or equal 
to 1 Gyr. Here, stellar ages are given as the age since the onset of star formation.
COS-207144 and COS-71929 have best-fit ages comparable to the
 age the universe at their redshift, whereas COS-75355 and COS-90676 have
 best-fit stellar ages a factor of 2-3 younger than the age of the universe at their 
redshift. The star-forming galaxy COS-37207 is 0.5 dex younger that the age of the 
universe at z=2.1. We note however that the stellar age is typically one of the most
 uncertain stellar population properties \citep{muzzin2009}.

Given the analysis presented in \autoref{Appendix} and summarized in 
\autoref{parameterdegeneracies} the metallicity is not constrained due to 
degeneracies in the fitted stellar population synthesis parameters.

\subsection{SPS Parameter Degeneracies}\label{parameterdegeneracies}
Because much of this analysis is based upon conclusions reached through the
use of SPS modeling, it is imperative to 
examine the degeneracies between the derived parameters of our SPS models.

In \autoref{Appendix}, we plot the correlations between EAZY estimates of 
redshift, FAST estimates of
stellar mass, SFR, dust attenuation, and population age for the
Monte Carlo simulations of both the photometry only and the photometry + spectrum. 
We use a Gaussian kernel density estimate in order to
visualize the distribution of the output parameters.

As expected, the FAST estimates increase in precision the most when a 
spectroscopic redshift is present. Though COS-37207 does not have a spectroscopic
redshift, its best-fit redshift shifted signficantly with the addition of the spectrum (see
\autoref{allspec}); the error on the FAST parameter estimates decreased 
slightly with the addition of the spectrum, but
the best-fit FAST parameter estimates remain relatively
unchanged.

As the only vigorously star-forming galaxy in the sample, it is of interest
to note that there exists a strong correlation between dust attenuation and
SFR in the Monte Carlo distribution of COS-37207. This correlation is not found in
any of the other galaxies' best-fit parameter distributions. This 
finding does not change the classification of COS-37207 as dusty and
star-forming, i.e., even at the lowest dust attenuation and SFR produced by the simulation, 
COS-37207 would be classified as dusty and star-forming. However, it reaffirms that
low-resolution spectra are not sufficient to break the degeneracy between SFR and
dust attenuation \citep{forsterschreiber2004, muzzin2009}.

As expected from other similar studies \citep[e.g., ][]{worthey1994,muzzin2009,lopezfernandez2016}, 
we find several other correlations between derived stellar population properties.
Stellar population age, dust attenuation, and metallicity all redden the
overall galaxy SED, and in the absence of prior knowledge 
that can break degeneracies \citep{conroy2012}, it is expected that there
will be anti-correlations between these parameters as derived from SED modeling.
For all of galaxies in our
sample, we see the expected negative correlation between stellar population
age and dust attenuation to some degree. The age-extinction
correlation is especially prominent in COS-207144 and COS-75355. This is 
likely because the addition of the spectrum (and spectroscopic redshift,
in the case of COS-75355) eliminated several competing parameter sets that
produced discrete populations within the Monte Carlo distribution, as can be seen
in the third panel of the left column of \autoref{75355attr}.

\begin{deluxetable*}{llllll}
\tablewidth{\linewidth}
\tablecaption{FAST Best-fit Parameters \label{fastoutput}}
\tablehead{
  \colhead{id} & \colhead{207144} & 
  \colhead{37207} & \colhead{71929} & 
  \colhead{75355} & \colhead{90676} }
\startdata
$\text{A}_\text{V}$ (mag) & $0.00^{+1.10}_{-0.00}$ & $1.70^{+0.10}_{-0.40}$ & $1.30^{+0.10}_{-0.40}$ & $0.30^{+0.10}_{-0.10}$ & $0.50^{+0.30}_{-0.50}$ \\
log (Age [yr]) & $9.50^{+0.00}_{-0.60}$ & $9.00^{+0.40}_{-0.00}$ & $9.50^{+0.00}_{-0.10}$ & $9.00^{+0.10}_{-0.10}$ & $9.10^{+0.20}_{-0.40}$ \\  
log ($\text{M}_\star/\text{M}_\odot$) & $11.46^{+0.00}_{-0.26}$ & $11.67^{+0.12}_{-0.01}$ & $11.87^{+0.00}_{-0.12}$ & $11.53^{+0.05}_{-0.02}$ & $11.64^{+0.06}_{-0.16}$  \\
log (SFR [$\text{M}_\odot~{\rm yr}^{-1}$]) & $-0.18^{+0.28}_{-1.82}$ & $2.53^{+0.12}_{-0.46}$ & $0.94^{+0.02}_{-0.67}$ & $-5.41^{+5.80}_{-5.80}$ & $-0.54^{+1.12}_{-1.46}$  \\
log (sSFR [${\rm yr}^{-1}$]) & $-11.64^{+0.37}_{-87.36}$ & $-9.14^{+0.10}_{-0.55}$ & $-10.93^{+0.09}_{-0.60}$ & $-16.93^{+5.75}_{0.00}$ & $-12.18^{+1.10}_{-86.82}$  \\
log ($\tau$ [yr]) & $8.50^{+-0.13}_{-1.50}$ & $8.50^{+0.50}_{0.00}$ & $8.60^{+0.00}_{-0.10}$ & $7.60^{+0.40}_{0.00}$ & $8.00^{+0.00}_{-1.00}$  \\
\textit{Z} & $0.03^{+0.00}_{-0.02}$ & $0.01^{+0.00}_{-0.01}$ & $0.00^{+0.00}_{0.00}$ & $0.01^{+0.02}_{0.00}$ & $0.00^{+0.03}_{-0.00}$  \\
$z_\text{peak}$ or $z_\text{spec}$ & $1.89^{+0.08}_{-0.06}$ & $2.11^{+0.01}_{-0.04}$ & $2.0000^{+0.0055}_{-0.0055}$ & $1.8220^{+0.0006}_{-0.0006}$ & $2.61^{+0.05}_{-0.08}$  \\
$q_\text{sf}$ & $0.993\pm0.005$ & $0.276^{+0.529}_{-0.276}$ & $0.963\pm0.028$ & $1.0^{+0.0}_{-0.008}$ & $0.999^{+0.001}_{-0.008}$ \\
\enddata 
\end{deluxetable*}

\subsection{$\text{M}_\star$ and SFR: Comparison with Independent Estimates}

In order to assess the accuracy of our previously-derived stellar population properties,
we compared our results to quiescent/star-forming classifications using the
rest-frame U-V versus V-J color-color diagram
 \citep[hereafter UVJ diagram; following, e.g.][]{whitaker2011,muzzin2013b,marchesini2014,whitaker2015,martis2016},
and SFRs derived from the UV and infrared (IR), as provided by \cite{muzzin2013}.

\begin{figure}
\centering
\includegraphics[width=\linewidth]{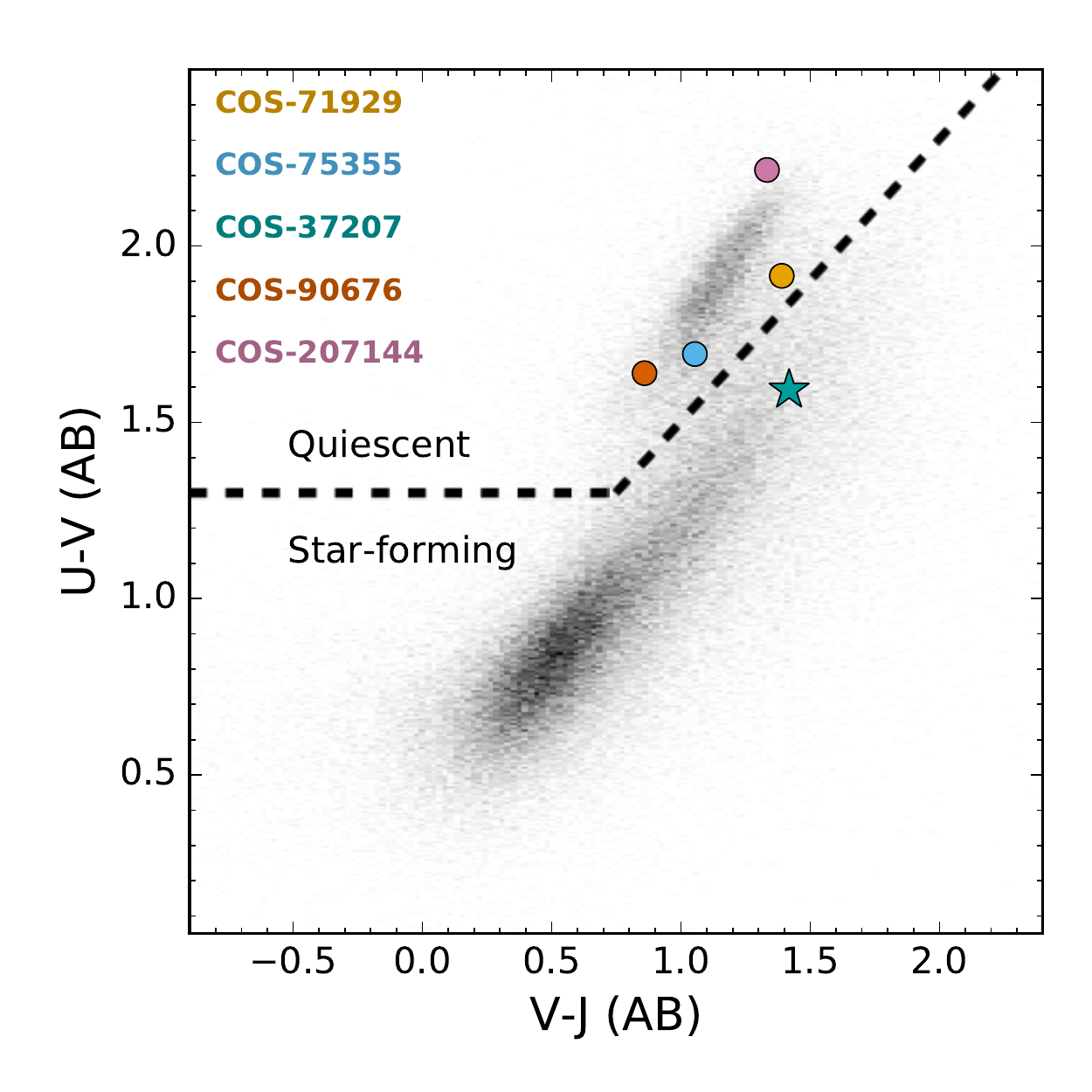}
\caption{
The UVJ diagram for the full UltraVISTA DR1 (grayscale) and the present sample (colored
 points, as labelled). The black dashed lines separate quiescent 
galaxies (\textit{top left}) 
from star-forming (\textit{right}). 
Given the position of COS-37207 in the UVJ diagram, 
this galaxy is characterized as dusty star-forming, in agreement with the results of the SED modeling.
The boundaries shown follow those presented in \cite{whitaker2015} and \cite{martis2016}
}
\label{uvj}
\end{figure}

\autoref{uvj} shows the targeted sample in the UVJ diagram using their original photometric
redshifts from \cite{muzzin2013}. 
Also plotted is the overall galaxy population in a stellar mass complete sample at the same 
redshift range as obtained using the UltraVISTA DR1 catalog of \cite{muzzin2013}. \autoref{uvj} 
shows that the location of the targeted galaxies in the UVJ diagram is consistent with the 
characterization of their stellar population properties as derived from FAST. In particular, we 
note that COS-37207 is in the region of the UVJ diagram typically populated by dusty 
star-forming galaxies \citep[e.g., ][]{martis2016}, while COS-71929 is in an intermediate zone between the 
dusty star-forming and quiescent galaxies. 

We checked the SED star formation rates for our sample against 
star formation rates derived using L$_{\rm IR}$ + L$_{\rm  UV}$ measurements
based on the UltraVISTA catalog.
We determine the rest-frame UV flux following \cite{muzzin2013}, estimating
$L_{2800}$ by integrating the best-fit template generated by EAZY \citep{brammer2008}
from 2600--2950~\AA.

Infrared luminosity ($L_\text{IR}$) was estimated using the $24~\mu m$
emission alone; we caution that this approach can produce individual measurements of $L_\text{IR}$ which
are uncertain up to several factors. A full explanation of the method to determine
$L_\text{IR}$ for the sample at hand can be found in \cite{wuyts2008}; briefly,
$L_\text{IR}$ is computed for the \cite{dale2002} infrared SEDs of star-forming galaxies 
at several heating levels of the interstellar environment. The best estimate of $L_\text{IR}$
is then taken to be the log-average of the set of $L_\text{IR}$ estimates for the
template across the range of heating levels sampled.

We calculate IR+UV SFRs following \cite{kennicutt1998} 
using the calibration of \cite{whitaker2014}, i.e.
$\text{SFR}_{\rm IR+UV} = 1.09 \times 10^{-10} ( {\rm L}_{\rm IR}+2.2 {\rm L}_{\rm UV}) L_\odot$.
\cite{bell2005} presents a near-identical calibration of $\text{SFR}_{\rm IR+UV}$,
which would yield the identical values of $\text{SFR}_{\rm IR+UV}$, within errors.
 The upper limits in the
IR+UV star formation rates indicate sources for which the MIPS $24 \mu m$ 
flux has a signal-to-noise ratio ${\rm S/N}<3$, and the $3\sigma$ upper limit
is adopted to estimate $L_{\rm IR}$. 

\begin{figure}
\includegraphics[width=\linewidth]{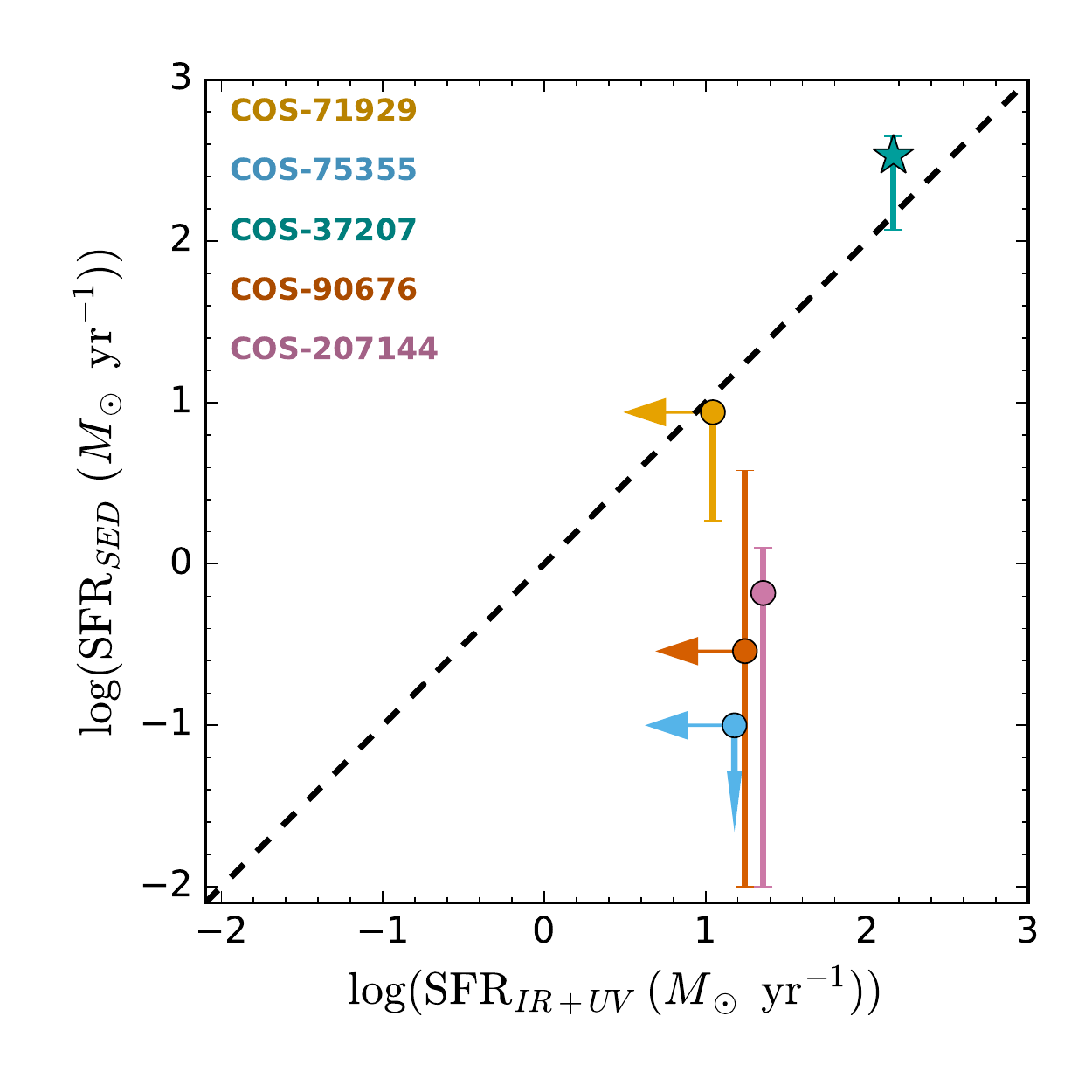}
\caption{
  Comparison between the SFR estimated from SED modeling and the SFR from 
  the combination of UV and IR measurements. Upper limits are given for the UV+IR SFR 
  when there is a non-detection in the MIPS 24$\mu m$ catalog photometry.
  With the exception of COS-207144, the galaxies in the sample at hand 
  have $\text{SFR}_\text{SED}$ estimates that are consistent 
  with the measurement of $\text{SFR}_{\text{UV}+\text{IR}}$ from \cite{muzzin2013}. 
  The black dashed line shows the 1:1 relation.
}
\label{sfrcompare}
\end{figure}

As shown in \autoref{sfrcompare}, for four out of the 
five galaxies in our sample, the SFR estimated from the SED and IR+UV are 
consistent with each other. Given that this includes three upper limits, however,
this statement demonstrates only that the SED SFR estimates are not 
significantly higher than expected.
COS-207144 is the only galaxy in our sample with a SFR derived from 
UV+IR much larger than the SFR from SED modeling. This result may be due
to a difference in the UV and IR properties of the two galaxies which make up
the source.

\section{Discussion and Conclusions}\label{discussion_and_conclusions}
In this work, we have presented NIR spectra of 5 ultra-massive galaxies in the
redshift range of $1.7<z<2.7$. We determined a spectroscopic redshift for
one of these galaxies based upon the presence of the \cahk{} absorption 
features in its spectrum, and estimated continuum redshifts for the
remaining galaxies in our sample using information from both  
broad- and medium-band UV-to-8~$\mu m$ photometry
and the aforementioned spectra. The stellar 
population parameters of the galaxies were also estimated, confirming the 
ultra-massive status of these galaxies.

Out of the five galaxies, only one -- COS-37207 -- is found to be star-forming,
with SFR~$\approx 340~\text{M}_\odot~yr^{-1}$.
This galaxy is also found to have the highest level of dust attenuation,
with $A_V$~$\approx1.70$~mag. 
A second galaxy, COS-71929, possesses properties that are intermediate with
respect to the quiescent and star-forming populations at this redshift.
The galaxy shows residual star formation activity, 
i.e., SFR$\approx$9~M${_\odot}$~yr$^{-1}$ and elevated dust obscuration, 
i.e., A$_{\rm V} \approx$1.3 mag, and
sits well below the main sequence of star forming galaxies at 
the redshift range of interest.
The remaining galaxies in the sample are
classified as quiescent with substantially less dust extinction by both SED 
modeling and UVJ diagram classification. 
This finding is in agreement with  
refined evolutionary path for the formation of local ultra-massive galaxies recently
presented by \cite{marchesini2014}. Though
the progenitors of today’s ultra-massive galaxies 
(i.e., $\text{log}~(\text{M}_\star / \text{M}_\odot)~11.8$ at $z \sim 0$) were massive, 
highly obscured, dusty star-forming galaxies at $2.5<z<3$, we expect the sample at hand
to quench at earlier times on average, as the galaxies
in the present sample have significantly higher stellar masses than the
majority of the galaxies from \cite{marchesini2014}. 
Though the present sample is too small to estimate population 
characteristics, if we consider COS-37207 to be the only
star-forming galaxy (grouping COS-71929, given its suppressed star formation,
with the quiescent galaxies in the sample)
we obtain a quiescent fraction of 0.80 for such galaxies 
($11.45<\text{log}~(\text{M}_\star / \text{M}_\odot)<11.9$) at $1.7<z<2.7$. 

There is not significant evidence that this quiescent fraction is systematically
larger than the fraction of $\sim 0.40$ estimated by \cite{marchesini2014} for the population 
of the progenitors of today's ultra-massive galaxies at $1.5< z<2.5$. \cite{marchesini2014}
found a quiescent fraction of $\approx0.40$ for their sample at $1.5< z < 2.5$, which covered a 
range of stellar masses of $11.2\la \text{log}~(\text{M}_\star / \text{M}_\odot) \la 11.5$.
Similarly, one out of the three galaxies with stellar masses 
$\text{log}~(\text{M}_\star / \text{M}_\odot)\gtrsim11.4$
in \cite{belli2014} was found to be star-forming at $2.<z<2.5$, and \cite{muzzin2013b}
estimates that the quiescent fraction should meet or exceed 0.50 at 
$\text{log}~(\text{M}_\star / \text{M}_\odot)\approx 11.3$ for the redshift range
considered in this work.

Assuming a true quiescent fraction of 0.40, there is a significant ($\approx 8\%$) chance
that a sample of five galaxies will contain at least four quiescent galaxies if the
sample were drawn without bias for the same distribution. 
However, given previous evidence that more massive galaxies enter quiescence at
earlier times relative to lower mass galaxies
\citep[see, e.g., ][]{brammer2011,muzzin2013b}, we find that our data are consistent
with these claims. 
However, in order to probe the true mass dependence of the 
quiescent fraction at a given redshift of such
ultra-massive galaxies a larger sample is needed in order
to properly marginalize over confounding variables, and
to suppress the uncertainty due to low number statistics.

In order to better characterize the population of 
ultra-massive galaxies in the early universe,
we must increase the number of identified ultra-massive galaxies and
construct a larger number of well-sampled SEDs from which to derive stellar
population property estimates and photometric redshifts. 
The fact that the inclusion of our NIR spectra, in the abscence of a
spectroscopic redshift, do not greatly affect the
stellar masses and redshifts of the galaxies in our sample implies that 
such galaxies may be reliably identified using photometry alone, in
agreement with \cite{muzzin2009}.
The NEWFIRM Medium-Band Survey II (NMBS-II) will identify these
ultra-massive galaxies across 5.2 ${\rm deg}^{2}$ of the sky (the
UltraVISTA DR1 catalog, upon which this work is based, spans 1.62 square
degrees), and will provide medium-band photometry to produce well-sampled
SEDs. Moreover, the VISTA VIDEO survey \citep{jarvis2013} includes
oft-visited areas of the sky and covers nearly 12 ${\rm deg}^{2}$, giving it
both the wavelength range and spatial coverage to robustly identify many
new ultra-massive galaxies.

Deeper spectroscopy will also greatly aid the study of this sample -- the 
galaxies at hand are bright enough that deep spectroscopy is feasible from the
ground, given an 8-10m class telescope. Though our detection of absorption
features in COS-71929 was not high enough S/N to 
perform a detailed measurement of the velocity dispersion, 
preliminary estimates indicate a substantial velocity
dispersion. Such measurements would also be 
greatly helpful in constraining the systematics of SED modeling by providing
independent measurements of output parameters.

\acknowledgements
\textit{Acknowledgments.} EKF acknowledges the support of the Tufts University Summer Scholar Program. DM acknowledges the support of the Research Corporation for Science Advancement’s Cottrell Scholarship and from Tufts University Mellon Research Fellowship in Arts and Sciences. ZCM gratefully acknowledges support from the John F. Burlingame and the Kathryn McCarthy Graduate Fellowships in Physics at Tufts University. DM and ZCM acknowledge support from the program HST-GO-12990, provided by NASA through a grant from the Space Telescope Science Institute, which is operated by the Association of Universities for Research in Astronomy, Incorporated, under the NASA contract NAS5-26555, and support from the National Aeronautics and Space Administration under Grant NNX13AH38G issued through the 12-ADAP12-0020 program.


KEW gratefully acknowledge support by NASA through Hubble Fellowship grant\#HF2-51368 awarded by the Space Telescope Science Institute, which is operated by the Association of Universities for Research in Astronomy, Inc., for NASA.

GHR acknowledges the support of NASA grant HST-GO-12590.011-A, NSF grants 1211358 and 1517815, the support of an ESO visiting fellowship, and the hospitality of the Max Planck Institute for Astronomy, the Max Planck Institue for Extraterrestrial Physics, and the Hamburg Observatory.  GHR also acknowledges the support of a Alexander von Humboldt Foundation Fellowship for experienced researchers.

Based on data products from observations made with ESO Telescopes at the La Silla Paranal Observatory under ESO programme ID 179.A-2005 produced by TERAPIX and the Cambridge Astronomy Survey Unit on behalf of the UltraVISTA consortium. 
Based on observations obtained at the Gemini Observatory, which is operated by the Association of Universities for Research in Astronomy, Inc., under a cooperative agreement with the NSF on behalf of the Gemini partnership: the National Science Foundation (United States), the National Research Council (Canada), CONICYT (Chile), Ministerio de Ciencia, Tecnolog\'{i}a e Innovaci\'{o}n Productiva (Argentina), and Minist\'{e}rio da Ci\^{e}ncia, Tecnologia e Inova\c{c}\~{a}o (Brazil). 
This paper includes data gathered with the 6.5 meter Magellan Telescopes located at Las Campanas Observatory, Chile. 

This work made used of the products from the 3D-HST Treasury Program (GO 12177 and 12328) with the NASA/ESA HST, which is operated by the Association of Universities for Research in Astronomy, Inc., under NASA contract NAS5-26555.

\bibliography{paper}
\clearpage

\begin{appendices}
\section{Monte Carlo Parameter Correlations}\label{Appendix}
Here we present the results of the Monte Carlo simulations run to constrain the error on our 
stellar population property estimates. For each run, fluxes for both the spectroscopy and photometry
are drawn from a normal distribution characterized by the error on the original measurement.
Each instance is run with a fixed redshift drawn according to the best-fit EAZY redshift PDF, as
shown in \autoref{allspec}.

\begin{figure*}
\centering
\includegraphics[width=\linewidth]{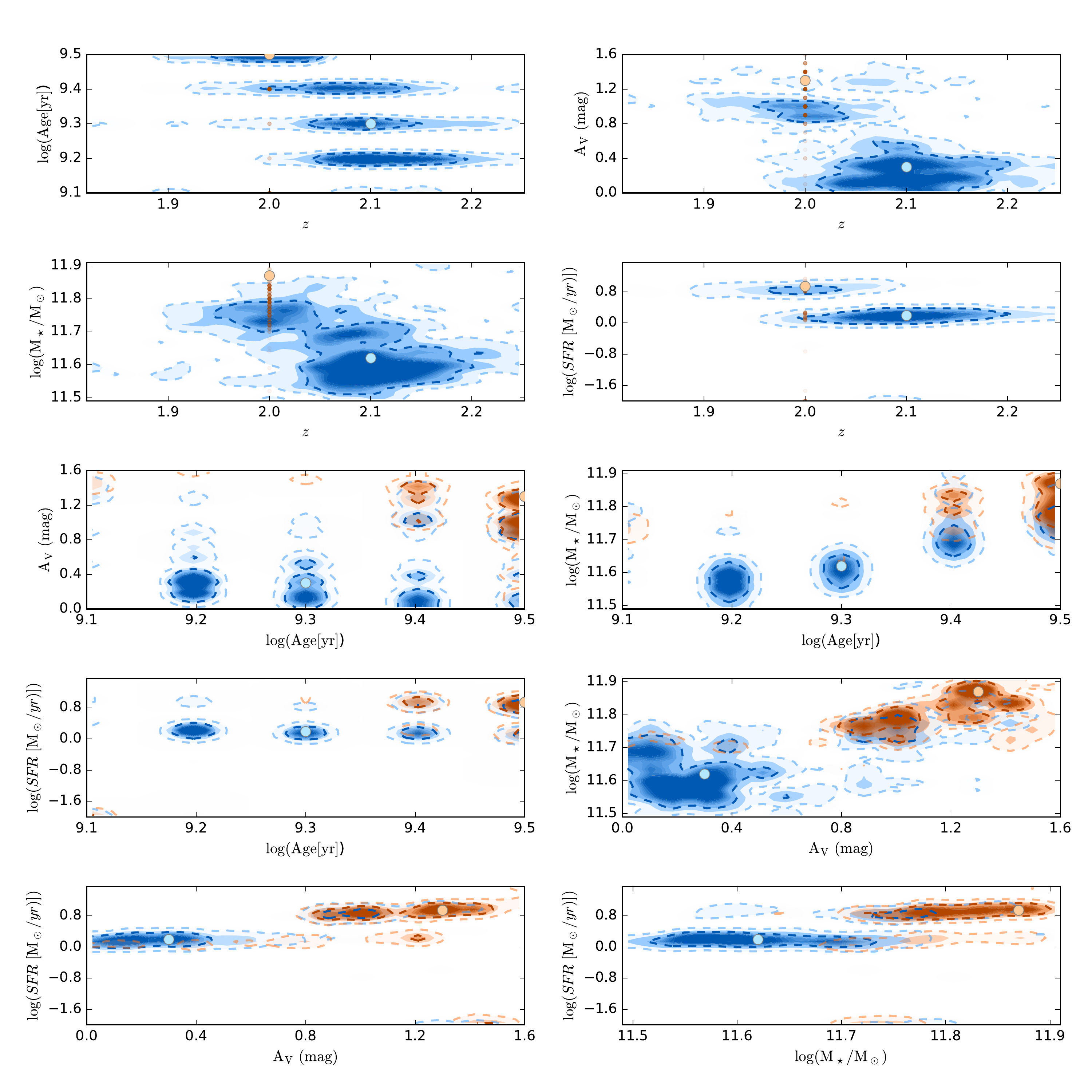}
\caption{Distributions of Monte Carlo best-fits FAST estimates 
for COS-71929 of
redshift, stellar population age, 
dust attenuation, stellar mass, and star formation rate. 
Monte Carlo distributions are shown for the case when FAST is
run with UltraVISTA catalog photometry only (blue) and with the 
spectrum in conjunction with the photometry (orange). Dotted contours show the
68\% confidence (dashed inner, darker curve) and 95\% confidence (dashed outer,
lighter curve) contours. The blue and orange points represent the best-fit 
FAST estimate from the original (unperturbed) spectrum for the photometry-only
and photometry + spectrum fit, respectively. }
\label{71929attr}
\end{figure*}

\begin{figure*}
\centering
\includegraphics[width=\linewidth]{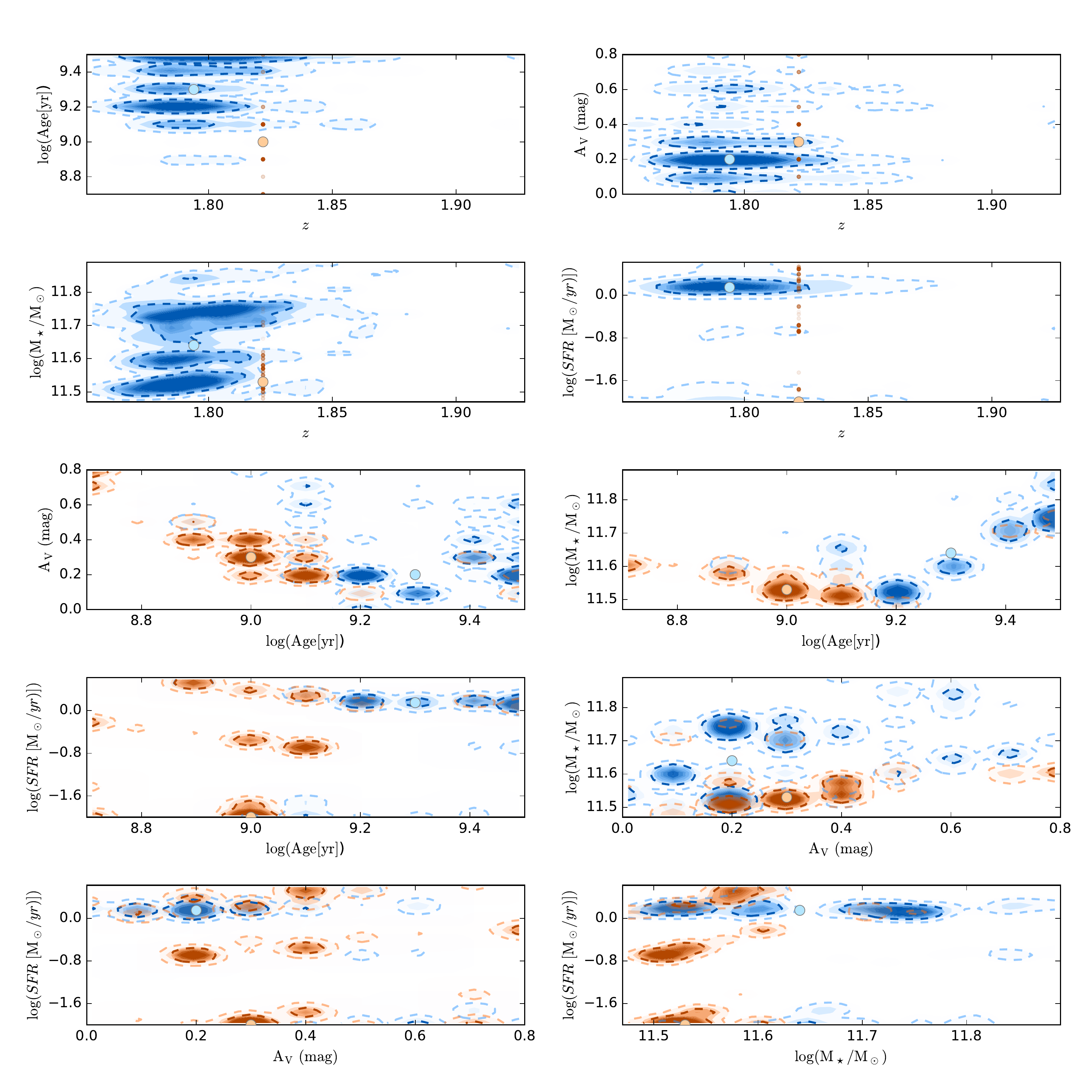}
\caption{Same as \autoref{71929attr} for COS-75355. }
\label{75355attr}
\end{figure*}

\begin{figure*}
\centering
\includegraphics[width=\linewidth]{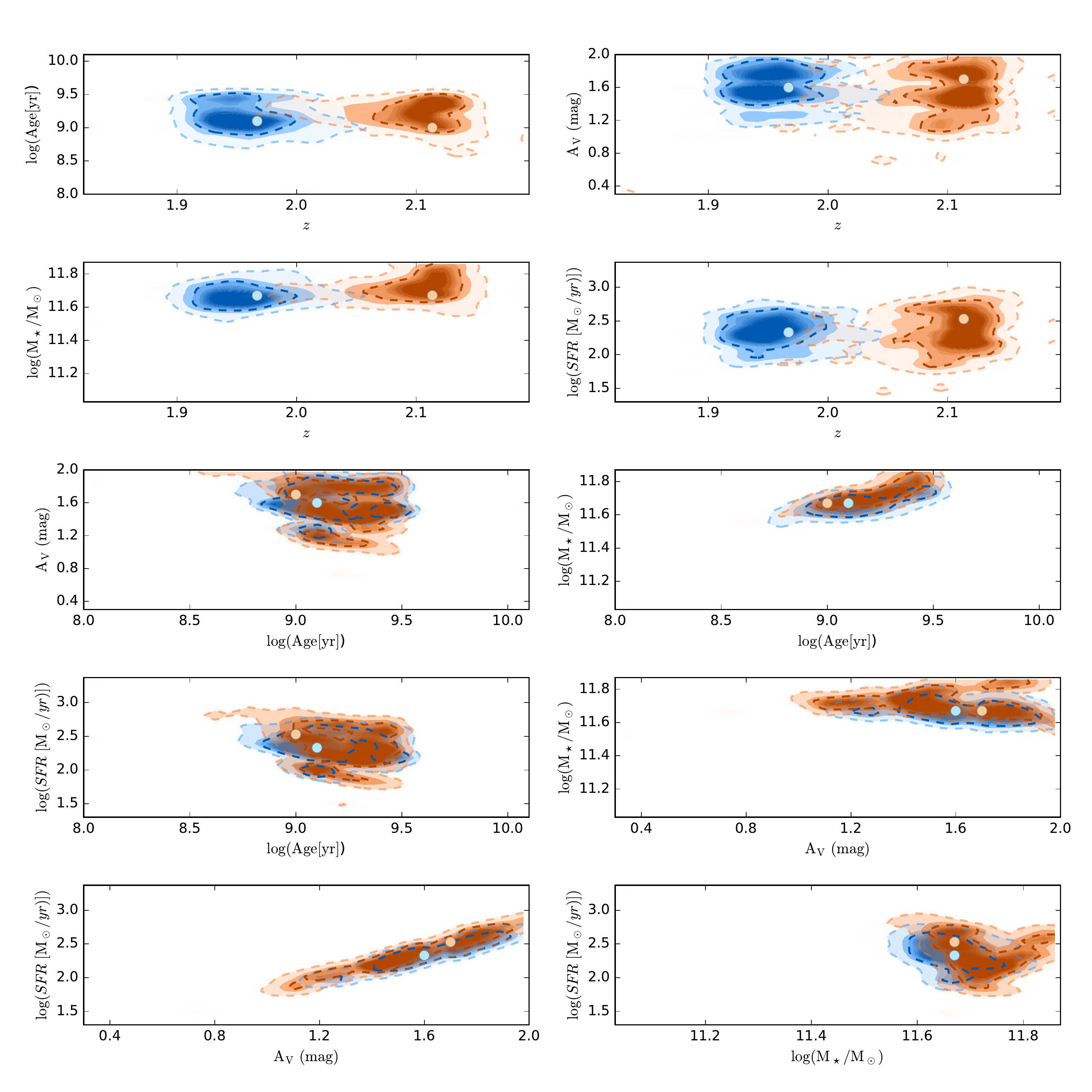}
\caption{Same as \autoref{71929attr} for COS-37207.}
\label{37207attr}
\end{figure*}

\begin{figure*}
\centering
\includegraphics[width=\linewidth]{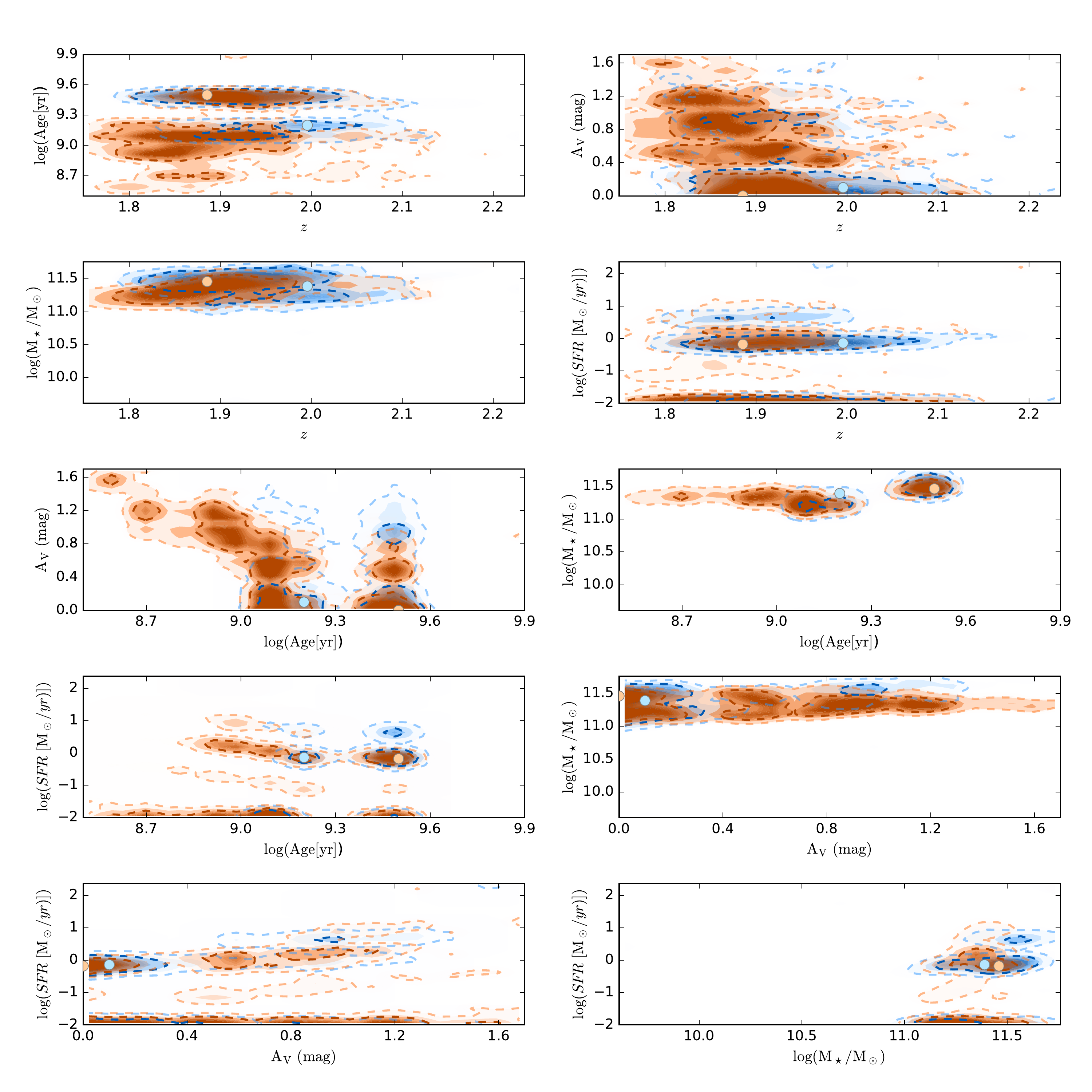}
\caption{Same as \autoref{71929attr} for COS-207144.}
\label{207144attr}
\end{figure*}

\begin{figure*}
\centering
\includegraphics[width=\linewidth]{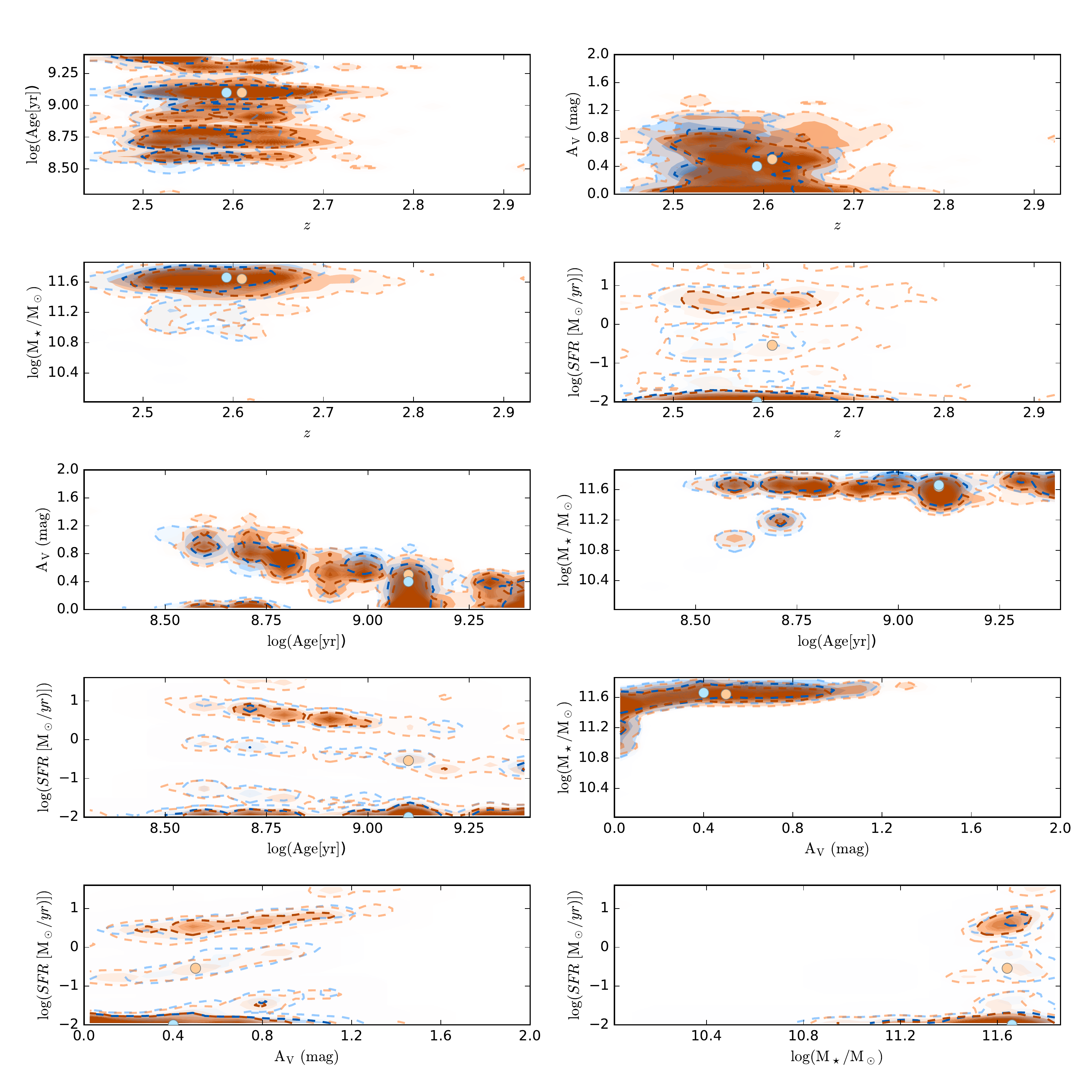}
\caption{Same as \autoref{71929attr} for COS-90676.}
\label{90676attr}
\end{figure*}

\end{appendices}

\end{document}